\documentclass[10pt, sigconf]{acmart}
\usepackage{amsmath}
\usepackage{hyperref}
\usepackage{enumitem}
\usepackage{listings}
\usepackage{cleveref}
\usepackage{booktabs} %
\usepackage{xspace}
\usepackage{xcolor}
\usepackage{microtype}
\usepackage{subcaption}
\usepackage{balance}

\usepackage[noend]{algpseudocode}
\usepackage{algorithmicx,algorithm}

\definecolor{darkgreen}{rgb}{0.15,0.55,0.15}
\definecolor{darkblue}{rgb}{0.1,0.1,0.5}
\definecolor{blue}{rgb}{0.19,0.48,0.9}
\definecolor{darkgreen}{rgb}{0.15,0.55,0.15}
\definecolor{mred}{rgb}{.80,.12,.30}
\definecolor{grey}{rgb}{0.5,0.5,0.5}
\definecolor{Purple}{rgb}{.75,0,.85}
\definecolor{light-gray}{gray}{0.95}
\definecolor{mid-gray}{gray}{0.85}
\definecolor{darkred}{rgb}{0.7,0.25,0.25}
\newcommand{\red}[1]{\textcolor{red}{#1}}

\newcommand{\highlight}[1]{{\smaller\colorbox{yellow}{\textbf{\texttt{\red{#1}}}}}}

\newcommand{\eat}[1]{}

\newcommand{\stitle}[1]{\vspace{2pt}\noindent\textbf{#1}}

\newlength{\listingindent}                %
\usepackage[LGRgreek]{mathastext}

\newtheorem{problem}{Problem Definition}

\DeclareMathOperator*{\argmin}{arg\,min}

\newif\ifnotes
\notestrue

\copyrightyear{2019} 
\acmYear{2019} 
\setcopyright{acmlicensed}
\acmConference[SIGMOD '19] {2019 International Conference on Management of Data}{June 30--July 5, 2019}{Amsterdam, Netherlands}
\acmBooktitle{2019 International Conference on Management of Data (SIGMOD '19), June 30--July 5, 2019, Amsterdam, Netherlands}
\acmPrice{15.00}
\acmDOI{https://doi.org/10.1145/3299869.3319872}
\acmISBN{978-1-4503-5643-5/19/06}

\newcommand{\review}[1]{#1}

\newcommand{\diff}{\texttt{$\delta$}\xspace}
\newcommand{\difftable}{\texttt{diffs}\xspace}
\newcommand{\sys}{Precision Interfaces\xspace}

\begin{document}

\author{Qianrui Zhang}
\affiliation{\institution{Tsinghua University}}
\email{zqr15@mails.tsinghua.edu.cn}

\author{Haoci Zhang}
\affiliation{\institution{Columbia University}}
\email{hz2450@columbia.edu}

\author{Thibault Sellam}
\affiliation{\institution{Columbia University}}
\email{sellam@cs.columbia.edu}

\author{Eugene Wu}
\affiliation{\institution{Columbia University}}
\email{ewu@cs.columbia.edu}

\setcounter{page}{1}
\setcounter{section}{0}
\begin{abstract}
  Interactive tools make data analysis more efficient and more accessible to end-users by hiding the underlying query complexity and exposing interactive widgets for the parts of the query that matter to the analysis.  However, creating custom tailored (i.e., {\it precise}) interfaces is very costly, and automated approaches are desirable.   We propose a syntactic approach that uses queries from an analysis to generate a tailored interface.  We model interface widgets as functions $I(q)\rightarrow q'$ that modify the current analysis query $q$, and interfaces as the set of queries that its widgets can express.   Our system, \emph{\sys}, analyzes structural changes between input queries from an analysis, and generates an output interface with widgets to express those changes.  Our experiments on the Sloan Digital Sky Survey query log suggest that \sys can generate useful interfaces for simple unanticipated tasks, and our optimizations can generate interfaces from logs of up to 10,000 queries in $\le10$s.

\end{abstract}

\title{Mining Precision Interfaces From Query Logs}
\maketitle

\section{Introduction}
\label{sec:intro}
\begin{figure}[h!]
    \centering
    \includegraphics[width=.7\columnwidth]{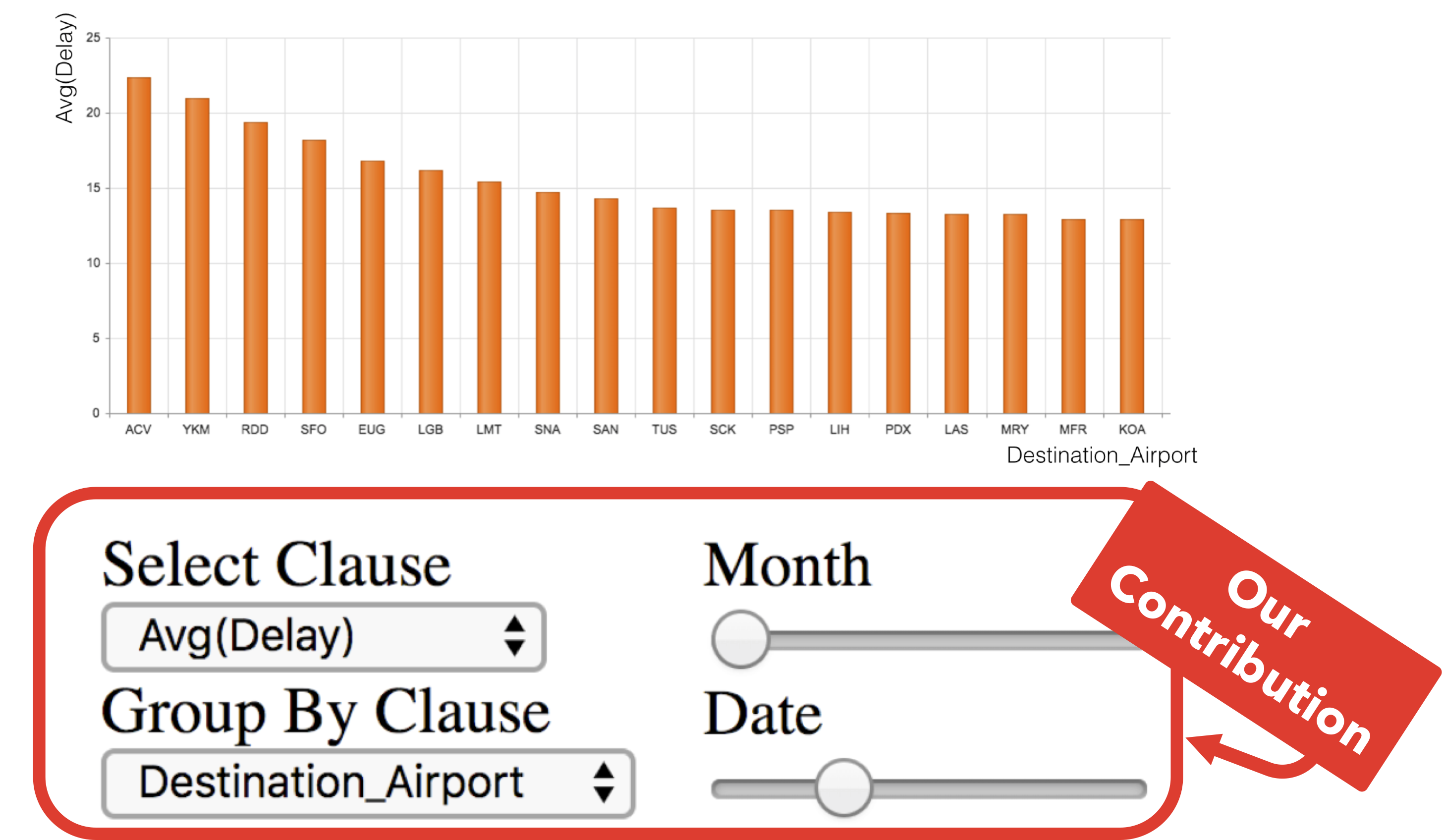}
    \caption{The OnTime flight delays dataset~\cite{ontime} contains 20 attrs.   Rather than an interface to express all analyses for all attrs, \sys generates an interactive interface specialized to the OLAP analysis queries in~\Cref{q:olap}. Users can pick from a small set of aggregation statistics, grouping attributes, and filtering conditions.  \sys focuses on generating the interactive components, and uses standard auto-visualization~\cite{mackinlay2007show} to render the results.}
    \label{fig:exampleinterface}
\end{figure}

\begin{figure*}[t]
\centering
  \begin{subfigure}[b]{.65\textwidth}
  \includegraphics[width=\textwidth]{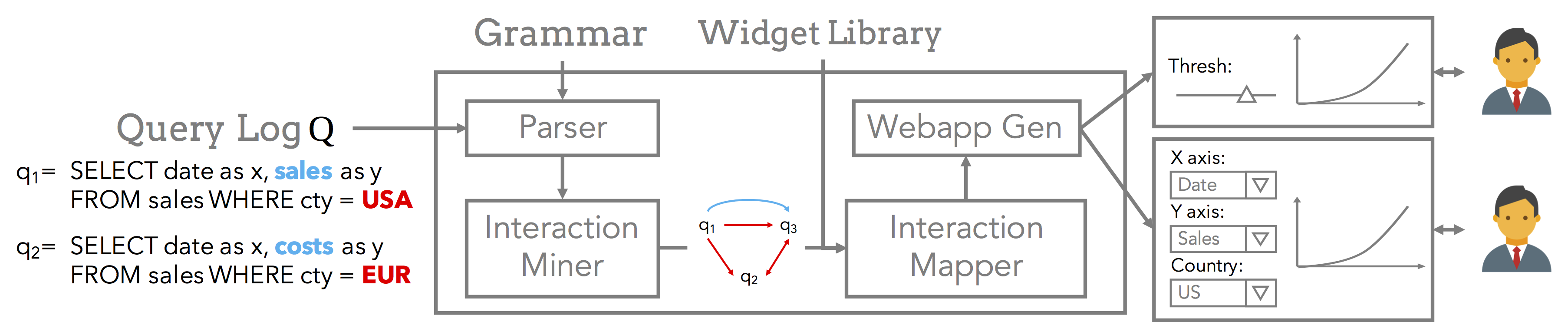}
  \caption{}
  \label{f:pipeline}
  \end{subfigure}
  \hfill
  \begin{subfigure}[b]{.325\textwidth}
  \centering
  \includegraphics[width=\textwidth]{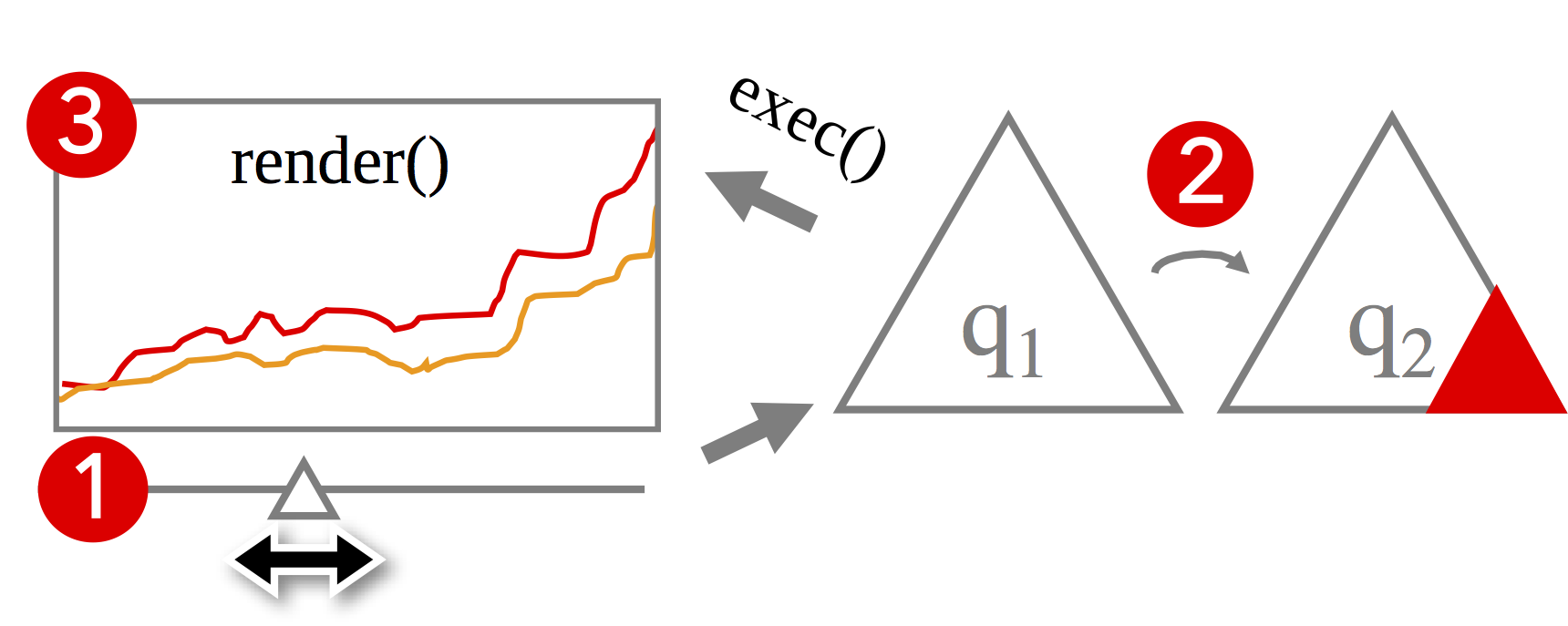}
  \caption{}
  \label{f:dummy}
  \end{subfigure}
  \caption{(a) \sys translates differences between queries into an {\it interaction graph} whose edges are mapped to interface widgets.  Each user's or analysis' queries creates a customized set of interactions.  (b) \protect\includegraphics[height=.9em]{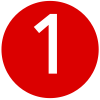}  User interaction changes \protect\includegraphics[height=.9em]{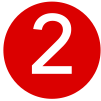}  AST of $q_1$ to $q_2$.  \protect\includegraphics[height=.9em]{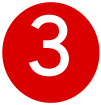} $exec(q_2)$ returns the query results and $render()$ visualizes it.}
  \label{f:arch}
\end{figure*}

End-users, businesses, and scientists increasingly rely on interactive visualization interfaces as their primary interface to analyze and monitor data.  These interfaces (e.g., \Cref{fig:exampleinterface},~\Cref{f:arch}) translate user manipulations of interactive widgets into queries whose results update the visualization.  Even for simple analyses, providing a simple interface helps hide the query complexity, lets users perform interactive analysis without programming, and is simpler for new users.  Although expert-designed interfaces are quite effective, identifying the relevant analysis and their queries, and developing the appropriate interface, is expensive.  It is desirable to automatically create interfaces specialized to an analysis.

A prominent approach is to use the database~\cite{jayapandian2008automated,Jayapandian2006AutomatingTD} to generate form-based interfaces to access and update the database.  However, what subset of a database's many tables, attributes, and possible queries are relevant for a specific analysis and should be expressible in the interface?  For example, the Sloan Digital Sky Survey (SDSS) database contains over 100 tables, and $2--60+$ attributes per table.  Further, most client's analyses in the SDSS query log involve small changes to simple queries.  For simple analyses, a generic interface designed to explore the entire database would be overly complex. 

A promising alternative is to use past queries to reduce interface complexity.  Query analysis has been used for query optimization~\cite{chaudhuri1998autoadmin}, query recommendation~\cite{eirinaki2014querie,Akbarnejad2010SQLQR,khoussainova2010snipsuggest}, web query analysis~\cite{hearst2009search,cai2016survey,chirigati2016knowledge}, and more.  However, its use to generate interactive interfaces has been less explored.  

Our primary insight is that an interface represents a set of queries: it renders the current query result, and widgets modify the query in user-understandable ways.  For instance, a slider changes numeric parameters, while a button may add a predicate or replace a subquery.  The set of queries that can be produced by all possible combinations of widget interactions can be viewed as the interface's expressive power.    Thus, we want to generate interfaces to express queries for a specific analysis.  Given existing and complementary work on automatic data visualization~\cite{mackinlay2007show,Moritz2018FormalizingVD} and interface layout~\cite{sears1993layout,myers2000past,zanden1990automatic}, \textbf{\textit{ the main technical challenge is to automatically map analysis queries to interactive widgets. } }

We present \sys, a system that automatically generates interactive interfaces from analysis queries (\Cref{fig:exampleinterface}).   We explore an extreme design point that primarily relies upon syntactic analysis of an input query log, and \emph{not} the database, to generate interfaces.  The system parses input queries into abstract syntax trees (ASTs), and maps differences between AST subtrees into interactive interface widgets.  This can be beneficial because the system is decoupled from the peculiarities of the language or dialect, and does not need access to the database.  Further, a general approach that supports SQL queries can potentially be extended to other query languages with minimal effort.  On the other hand, there are potential drawbacks from not leveraging query semantics, database schemas, and the actual data distributions, and we discuss them in~\Cref{s:assumptions}.

For these reasons, \textbf{\textit{our major contribution is a unified model that connects queries, query transformations, interactions, and interfaces}}.  We describe how interactions can be modeled as tree transformation functions that are mined from analysis queries, and how to map these interactions to a general widget library.   In addition:
\begin{enumerate}[leftmargin=*]
\item We develop several pragmatic optimizations that speed up the performance by multiple orders of magnitude, so that  \sys can generate interfaces for query logs containing up to $10k$ queries in $<10$ seconds. 
\item We highlight the strengths and weaknesses of \sys using 3 different query logs---a synthetic OLAP exploration log, the SDSS query log, and an open ended exploration log.  We show that the complexity of the generated interfaces scale with the variety and complexity of the query {\it transformations}, and are independent of the query complexity.  When the analysis queries change in structured, predictable ways, \sys only needs a few dozen query examples to generate interfaces that can express hundreds of subsequent queries in the same analysis.  \sys does not work as well when changes are unpredictable.
\item Our user study compares an interface generated using queries from four SDSS analysis tasks, against the SDSS search form interface.  We find that \sys can identify and automatically generate a task-relevant interactive interface with widgets for a task that required a ``write SQL'' fallback in the SDSS interface.  For that task, the interface is considerably faster and more accurate to use.  Overall, the interface was qualitatively preferred by all participants.
\end{enumerate}

\noindent This paper shows that it is possible to automatically generate sensible interactive analysis interfaces from query logs that contain recurring structural transformations.   There is still ample room to improve the quality of the output interfaces, and our experiments show the limitations of a purely syntactic approach when analyzing complex query logs with random queries or mixtures of queries from many users.   We describe the assumptions and limitations in~\Cref{s:assumptions}, and future directions in~\Cref{s:conclusion}

\section{Related Work}\label{s:related}

\stitle{User Interface generation:}
\review{Jayapandian et al. automate form-based record search and creation interfaces by analyzing database contents~\cite{jayapandian2008automated,Jagadish2007MakingDS,Jayapandian2006AutomatingTD}.  This work is complementary, in that they rely solely on the database schema and data, rather than the query log, and may generate overly complex forms even if individual analyses are simple.  } Our future work plans to incorporate data and query semantics.

The UI literature focuses on form layout and design, and relies on the developer to provide a high level specification of the tasks and data~\cite{puerta1994model,vanderdonckt1994automatic,nichols2004improving}.  The above works do not explicitly leverage query logs.  In contrast, we use example queries to synthesize {\it interactive} analysis-specific interfaces. 

\stitle{Log Mining:} Historically, query log mining has been used in the database literature to detect representative workloads for performance tuning~\cite{chaudhuri1998autoadmin,hellerstein2007architecture}, and in the web query literature~\cite{silvestri2009mining} to e.g., augment web search results~\cite{hearst2009search,Fourney2011CharacterizingTU}, make keyword suggestions~\cite{cai2016survey}, or enable exploration~\cite{chirigati2016knowledge}.  

SQL query analysis has also been used to support data exploration. \review{QueRIE~\cite{eirinaki2014querie,Akbarnejad2010SQLQR}, SnipSuggest~\cite{khoussainova2010snipsuggest}, SQB~\cite{Khoussainova2011SessionBasedBF} produce context-sensitive query suggestions and summaries of related queries from an existing query log by analyzing the log at the string level.  This work is complementary to ours, which can then map the recommended query fragments to interaction widgets.   Query steering~\cite{dimitriadou2014explore} uses a Markov model to produce new statements. VQIS~\cite{Christodoulakis2017VIQSVI}, leverage a log of templated queries along with semantic annotations of table attributes to provide richer recommendations; Yang et. al~\cite{Yang2007NuggetDI} develop a query similarity measure that takes results into account. The work on inferring query sessions~\cite{Khoussainova2009ACF} could help \sys disambiguate queries part of different analyses.} 
\sys summarizes the structural changes in query logs as interactive interfaces.

\stitle{Development Libraries and Dashboards:} Tools such as Sikuli~\cite{yeh2009sikuli} or Microsoft Access let non-technical users build their own interfaces. They improve upon lower-level libraries (e.g., Bootstrap) but still require programming and debugging. Reactive languages (e.g., Shiny~\cite{chang2015shiny}, EVE~\cite{eve}) still require programming and are limited to value changes rather than structural program changes.  \review{Similarly, dashboarding companies services (e.g., Metalab, Looker, etc) help visualize complex queries and provide widgets to change query parameters.  Our system goes beyond this by identifying and supporting more complex structural changes. }

\stitle{Visualization Recommendation:} Visualization recommendation tools such as Panoramic Data~\cite{zgraggen2014panoramicdata}, Zenvisage~\cite{siddiqui2016effortless} and Voyager~\cite{wongsuphasawat2016voyager} constitute a recent and complementary research direction. Those tools help recommend similar data to a given view, while \sys seeks to generate the exploration interface itself.  Further, \sys can leverage automatic visualization designers, such as ShowMe~\cite{mackinlay2007show}, Draco~\cite{Moritz2018FormalizingVD}, APT~\cite{mackinlay1986automating}, to interactively visualize query results in the generated interface.  

\stitle{Interface Redesign:}  Interface redesign adjusts the layout and/or selects alternative widgets based on the display size, modality~\cite{adaptive1993}, personalization~\cite{gajos2007automatically,gajos2010automatically,weld2003automatically}. Survey form redesign has also been used to reduce data-entry errors~\cite{kuangusher}. Those techniques are complementary to ours, which focuses on identifying and selecting task-specific interactions.

\stitle{Programming Languages:} \sys can be viewed as learning a domain specific language (DSL)~\cite{crespi1973use} that is expressed in the interaction domain.  Program synthesis seeks to construct programs that satisfy a high level logical description. For instance, Potter's Wheel~\cite{raman2001potter} and Foofah~\cite{jin2017foofah} build data transformation programs based on input and output examples. We target a different problem---\sys{} analyzes query logs, not input-output pairs, and it synthesizes interfaces.

\section{Motivation and Architecture} \label{sec:overview}



Interfaces are traditionally created by programmers or through a WYSIWYG application, so why mine interfaces from query logs?  The primary reason is that query logs encode the analyses that analysts {\it actually} perform, and therefore can be used to suggest candidate interfaces. Using logs as the system API is natural because they are generated by many sources.  Modern program execution engines (e.g., DBMSs, Spark, Jupyter, RStudio) already track program logs for recovery and debugging purposes, while explicit provenance meta-data systems are increasingly ubiquitous~\cite{ground,ives2008orchestra,muniswamy2006provenance,callahan2006vistrails}.  Further, any analysis or application using these systems (e.g., Tableau) will naturally collect logs.  

We now describe several motivating use cases, the system overview, and outline our assumptions.

\subsection{Use Cases}\label{s:usecases}

Custom designed interfaces will typically be much better than those generated by \sys.  However, we anticipate a number of compelling use cases when dedicated developers and UI designers are unavailable.  This is commonly the case for ``long-tail'' analyses where one or two users may often perform them, but there are not enough users to justify custom design efforts.

\stitle{Tailored dashboards:} An IOT startup (name anonymized) regularly performs tailored analyses for its customers. The engineers wrote a custom dashboard builder for simple queries, but it does not support complex statements (e.g., nested queries) nor analyses.   Therefore, the employees (including the Chief Scientist) spend considerable time writing queries. For each case, they retrieve a text file containing past customer queries, identify the statements that they need, customize and copy-paste them, and possibly update the file and check it into version control. A tool to build interfaces from queries would allow them to quickly set up expressive front-ends for each case and each customer.

\stitle{Interface Simplification: } Interfaces, such as the SDSS interface described in the user study (\Cref{s:userstudy}), or even the SQL language, are often designed to support a wide-range of use cases and tasks. This can be challenging for a new user to both understand the general interface and how to use it to accomplish a single task.  \sys is one approach to identify the queries specific to tasks that users perform in practice, and generate simpler interaction controls for them.  These interfaces can serve as ``fast-paths'' when the task is clear, but the user can fall back to existing interfaces for alternative tasks or to go ``off-script''.   In addition, the interfaces can serve as a starting point for designers.

\subsection{System Overview}\label{s:sysoverview}

Interfaces are largely composed from a common set of widgets used to change values, attributes, and queries. However, without guidance from a user, what widgets should be added to the interface, and what should those widgets do?  A benefit of analyzing queries is that they help narrow the space of allowable changes to consider.

We decompose the problem of generating interfaces from query logs into two sub-tasks: finding structural changes between queries, and mapping those changes to interactive widgets. In Figure~\ref{f:pipeline}, the user submits a query log $Q$ which is parsed using a lightly annotated grammar  (Section~\ref{s:model-progs}).  Each query is parsed into an abstract syntax tree (AST); the {\it Interaction Miner} aligns pairs of ASTs to identify the set of subtree differences that transform one query into the other.  

These differences form an {\it Interaction Graph}, where each query is a vertex and subtree transformations are edges.  Given a library of widget types (e.g., drop-downs, sliders, text-boxes), we instantiate widgets to express groups of edges.  For instance, a drop-down could let users select $USA$ or $EUR$, which changes a string literal node in the AST.  In contrast, a toggle button may directly replace the entire query's AST.  This problem is NP-hard, and we present a graph contraction-based heuristic that iteratively merges redundant widgets.  Once the appropriate widgets have been mapped, the system generates \review{an editable grid of widgets. The user can customize the widget labels, positioning, and parameters, and then ``compile'' the layout into an interactive web application.}  Different logs generated by, say, different users or different analyses, are processed separately and result in different precision interfaces.

Figure~\ref{f:dummy} depicts how a generated interface operates.  The visualization initially renders the output of $q_1$. The slider is mapped to a threshold parameter in a (potentially) complex aggregation query.  \includegraphics[height=.9em]{figures/c1} When the user interacts with the slider, the slider's current value is used to generate a new subtree (the red triangle).
\includegraphics[height=.9em]{figures/c2} The subtree replaces the existing subtree at the location of the threshold parameter, thus transforming $q_1$ to $q_2$.  
\includegraphics[height=.9em]{figures/c3} The $q_2$ AST is then executed by calling \texttt{exec($q_2$)} and \texttt{render()} is called to update the visualization with the new query's results.

\stitle{Challenges: } 
Real-life query logs may contain much variability, and it is not obvious how to map  arbitrary AST tree differences to different types of widgets automatically. 
In general, any widget can express anything: a button press could replace the current query with a random query, or a slider could map each slider position to an arbitrary query.  Without careful thought, it is easy to select overly complex, or dysfunctional combinations of widgets.

This leads to several technical challenges.
\textbf{Modeling:} what is a unified model of queries, interactions, interfaces and interactive widgets that is restricted enough for analysis but rich enough to support custom widgets and user preferences?  \textbf{Quality:} how to generate a compact set of widgets that expresses analyses represented in the log, and minimizes superfluous widgets?   \textbf{Runtime:} how to ensure fast runtime without affecting the output quality?

\subsection{Assumptions and Limitations}\label{s:assumptions}

Data-driven interface generation is deeply challenging.  There are many simplifying assumptions and limitations in this work, and considerable opportunity in this area.  

We do not assume deep semantic understanding about queries beyond near-universal features such as primitive data types.  Instead, we perform syntactic analysis on the abstract syntax tree (AST). \review{\Cref{s:model-progs} further describes our assumptions of the language grammar in more detail.} Further, we don't leverage side-information such as the database schema nor contents for two reasons.  First, there are many cases, such as working with a company, where we have access to query logs but not the data.   Second, our longer term goal is to support other query and scripting languages (e.g., SPARQL, Python, R), thus we do not want to rely on database access.  

There are limitations to a syntactic, change-based approach.  Notably,  we cannot distinguish syntactically different but semantically equivalent queries, nor identify syntactically correct but semantically incorrect queries (see experiment in \Cref{a:exp_precision}).  Leveraging the database contents~\cite{jayapandian2008automated} and query-specific semantics could help.

We assume two available functions \texttt{exec()} and \texttt{render()} that respectively execute a query AST and render the output.  \texttt{exec()} is called on any user interaction that changes the interface's current query, and \texttt{render()} either generates a simple visualization~\cite{mackinlay2007show,mackinlay1986automating} or renders a table.

We assume that there is no logical dependency between the entries in the log---for instance, that a result value from a previous query (e.g., ``\texttt{IBM}'') is used as a parameter of a subsequent query (e..g, \texttt{SELECT ... name =} ``\texttt{IBM}''), because it requires access to the database or query results.  We also assume that view or temporary table references have been expanded into sub-queries so that each query is self-contained.  

This work assumes that the query log contains queries from a single logical analysis, in that it exhibits recurring structural transformations that are predictable.  Our experiments in \Cref{s:exp_recall_multiclient} evaluate heterogeneous logs that combine queries from multiple user sessions.  Although \sys is capable of generating widgets to express the queries, the resulting interface is quite complex because the widgets not only need to express the individual analyses, but ways to translate between the analyses.  Preprocessing the query log by leveraging query meta-data (e.g., session IDs are automatically stored in DBMS query logs), modeling semantic distances between queries~\cite{Yang2007NuggetDI,Christodoulakis2017VIQSVI} to cluster similar queries, and removing ``anomalous'' queries are all promising approaches.  However, preprocessing should be sensitive to cases where anomalous queries are crucial for the analysis.

\section{Model and Problem Definition}\label{s:model}

Our goal is to define good mappings from query logs to interactive widgets in interfaces.  To do so, it is crucial to define a unified model of queries, query differences, interactions, and interfaces.   The key idea is that interactive widgets express query transformations that we mine from the query logs.

\subsection{Queries as Parse Trees}\label{s:model-progs}

Let $Q$ be a sequence of queries from a given analysis; we wish to express their recurring structural differences using interactive widgets.    This work does not leverage the semantics of the SQL query language, and instead analyzes each query $q_i$ as its parsed abstract syntax tree (AST)\footnote{We do this to more readily support different dialects, and in the future, different analysis languages.}.  We use the language grammar and a minimal set of grammar annotations to parse and interpret the ASTs.

\begin{figure}
\centering
\includegraphics[width=\columnwidth]{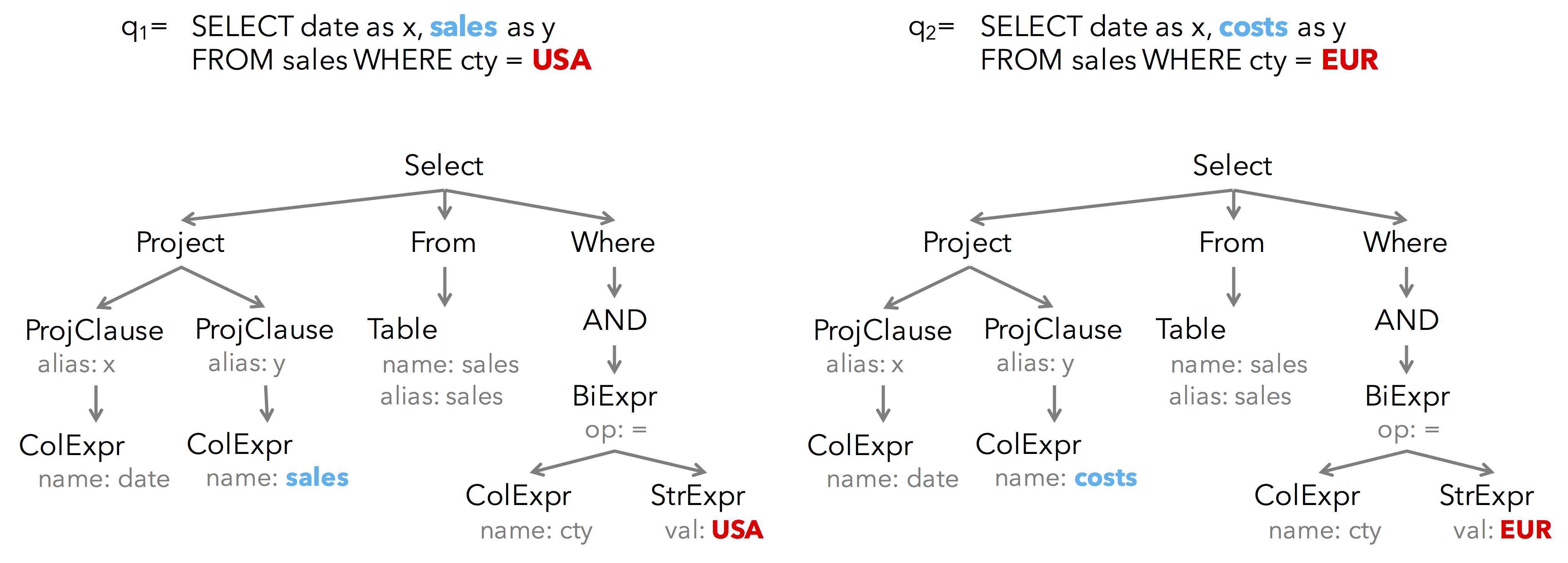}
\caption{\small Example ASTs for two SQL queries that differ in the second project clause (blue) and the constant in the equality predicate (red). }
\label{f:asts}
\end{figure}

Figure~\ref{f:asts} shows simplified ASTs for two example queries.
Each node consists of its type, a set of attribute-value pairs, and an ordered list of child nodes.  For instance, \texttt{cty=USA} has node type \texttt{BiExpr}, attribute-value pair \texttt{op:`='}, and two children for the left and right sub-expressions.  Its second child is a string literal node \texttt{StrExpr} with value \texttt{USA}.

\stitle{Assumptions: }  We assume that there is a mapping from some terminal node types to primitive data types (e.g., \texttt{StrExpr} maps to a string literal, \texttt{IntExpr} maps to an integer).  This is because some interactive widgets, such as numeric sliders, are typed.  Similarly, we assume knowledge of node types that represent collections of sub-expressions (e.g., for simplicity, we model \texttt{Project} as a collection of \texttt{ProjectClause} nodes).  This is because widgets such as checkboxes model a collection of options.  This mapping can be automatically identified based on common grammar idioms\footnote{SQLite grammar defines the list of output expressions \texttt{sel\_core} as a project clause (represented by the \texttt{sel\_result} non-terminal) followed by zero or more additional project clauses: \texttt{sel\_core = (sel\_result (whitespace comma sel\_result)*)}}, or manually annotated once per language/dialect.

\begin{table}\footnotesize
\hspace*{-.1in}
\begin{tabular}{ccclllc}
   \toprule
   & \textbf{$q1$}  & \textbf{$q2$}    & \textbf{$\pi$}& \textbf{$\tau_1$}& \textbf{$\tau_2$} &\textbf{$type$} \\
   \midrule
  $\delta_1$ & 1 & 2 & 0/1/0   & \texttt{ColExpr(sales)} & \texttt{ColExpr(costs)} & str \\
  $\delta_2$ & 1 & 2 & 2/0/0/1 & \texttt{StrExpr(USA)} & \texttt{StrExpr(EUR)} & str\\
  $\delta_3$ & 1 & 2 & 0/1 & \texttt{ProjClause} & \texttt{ProjClause} & tree \\
  $\delta_4$ & 1 & 2 & 2/0/0 & \texttt{BiExpr} & \texttt{BiExpr} & tree \\
   \bottomrule
\end{tabular}
\caption{\diff records in \difftable table for ASTs in Figure~\ref{f:asts}.}
\label{t:difftable}
\end{table}

\subsection{Interactions as Query Differences} \label{sec:diff}

\sys models query differences as subtree transformations between each pair of queries $q_i$ and $q_j$.  The goal is to map common subtree differences to interactive widgets in the interface.  We begin with an example:

\begin{example}\label{e:asts}\it
  Consider the ASTs in Figure~\ref{f:asts}.  A trivial transformation is to replace the root of $q_i$ with the entire AST of $q_j$.  For instance, a toggle button could simply replace one tree for the other, and vice versa.    A more fine-grained transformation would be to replace the minimally sized subtrees.  In the example, there are two such subtrees:  \texttt{ColExpr} from ``sales'' to ``costs'', and \texttt{StrExpr} from ``USA'' to ``EUR''.  Given these, their ancestor subtrees are naturally also valid transformations.  
\end{example}

A given subtree transformation between $q_i$ and $q_j$ is specified by a tuple $\diff_k = (\pi, \tau_i, \tau_j)$. $\pi$ specifies the path to the root of the subtree changes; $\tau_i$ and $\tau_j$ specify the subtrees rooted at $\pi$ in their respective queries.  Subtree additions and deletions are represented by setting $\tau_1$ or $\tau_2$ to \texttt{null}, respectively. 
Let \difftable model all transformations in the query log\footnote{Note that this table is logical, and need not be full materialized.}.  For example, Table~\ref{t:difftable} shows \difftable for the above example.

\begin{example}\it
  Consider the first row $\diff_1$ in Table~\ref{t:difftable}. Its path $\pi$ follows \texttt{PROJECT} (0/), to the second \texttt{ProjClause} (0/1/), to its only child (0/1/0).  It replaces the value of the column expression {\it sales} ($\tau_1$) with {\it costs} ($\tau_2$). The last column states that the change is between string literals.  
\end{example}

Finally, $\diff$ can be interpreted as a function $\diff(q) = q'$ that replaces the subtree rooted at $\pi$ in $q$.  Similarly, its inverse $\diff^{-1}(q') = q$ replaces $\tau_1$ at location $\pi$ in $q'$ to recover $q$.

\stitle{Interactions: } Interactions are the abstraction that connect query transformations with interactive widgets in the interface.   An interaction $t\subseteq\difftable$ models the set of differences needed to fully transform one query to another.  For instance, $t = \{\diff_1, \diff_2\}$ states that the changes described by the first two rows in Table~\ref{t:difftable} are sufficient to transform $q_1$ to $q_2$.  Specifically, $q_2 = t(q_1) = \diff_1(\diff_2(q_1))$.   Note that there can be many possible interactions between two pairs of queries.  For instance, $\{\diff_1, \diff_4\}$ is also sufficient to transform $q_1$ to $q_2$.

\stitle{Interaction Graph: }  \difftable describes the set of edges between queries in the input log.   We model it as an {\it interaction graph} $G=(V,E)$, where each query $q\in Q$ is a vertex $v$, and a directed edge $e = (q_i, q_j, t_k)$ is labeled with an interaction such that $t_k(q_i) = q_j$.  As implied above, there can be multiple labeled edges between any two queries, and each labeled edge corresponds to one or more \diff records.  This graph representation is useful when describing the interface mapping problem below.  

\stitle{Implementation:} To find subtree differences, we use a fast ordered tree matching algorithm~\cite{bille2005survey,thomas2001introduction} that preserves ancestor and left-to-right sibling relationships when matching nodes between the two trees. The algorithm first computes the pre-order traversal of both trees. It goes on to the next node if the current pair of nodes matches. When the algorithm finds a pair of nodes that cannot be mapped, it uses backtracking to return to the last pair of nodes that has already been mapped and tries to map them to some other candidate.  The algorithm has $O(\Pi_{i\in\{1,2\}} (T_i \times min(L_i,D_i)))$ complexity where $T_i,L_i,D_i$ are respectively the size, number of leaves, and the tree depth of the $i^{th}$ tree.

\subsection{Interaction Widgets}\label{s:model_widgets}

We model an interface $I$ as a set of interactive widgets.  For example, the interfaces in Figure~\ref{f:pipeline} consists of the rendered output (which we assume is provided by a \texttt{render()} method), along with three drop-down widgets. Each widget $w_i$ is an instance of the drop-down widget type $WT$, and is customized to change a specific part of the query.  The top widget changes the grouping attribute, the middle widget changes the aggregated attribute, and the bottom widget changes a predicate value.   Thus, despite all being the same type of widget, their effects on the query are different.  We now define widget types and widget instances.

\stitle{Widgets and Widget Types: } A given widget type is well suited for particular types of AST transformations.  For instance, numeric sliders are well suited to specify a value from a range of numbers, whereas a drop-down picks a value from a small set of options.  We model a widget type $WT = (r_{WT}, c_{WT}())$ as a combination of a constraint rule $r_{WT}$ and cost function $c_{WT}$.  \review{$WT$ is instantiated as a widget $w$ by specifying a path $w.\pi$ in the AST that the widget will modify using a subtree in the widget's domain $w.d$, where $w.d$ is initialized by a subset $w.\Delta\subseteq \difftable$.  The widget domain represents the set of allowable subtrees that a widget can express, whereas $w.\Delta$ is the subset of \difftable used to initialize the widget (see the slider example below).}

\begin{example}\it
  The domain of the top drop-down in Figure~\ref{f:dummy} may be the subtrees for the string literals {\it Date}, {\it Hour}, and {\it Week}.  Similarly, $\delta_1$ in Table~\ref{t:difftable} defines the domain $\{$\texttt{ColExpr(sales)}, \texttt{ColExpr(costs)}$\}$.  The domain need not explicitly enumerate a set of subtrees.  For instance, a slider's domain may be initialized with $w.\Delta$ containing the subtrees $\{1, 5, 100\}$, but its domain will be extrapolated as the range $[1, 100]$. In this way, it can express all values between $1$ and $100$, even though $w.\Delta$ only contained three subtrees.
\end{example}

\stitle{Widget Rule: } Rule $r_{WT}(w.d)$ is a function that checks whether $w.d$ satisfies the conditions to instantiate an instance of the widget type $WT_i$.  For instance, the slider widget type only accepts subtrees that represent number literals.  If $d$ contains any other type of subtree, then $r_i(w.d)$ returns false.  In general, $r()$ will enforce that the elements in a domain $d$ are all of a particular type.  In our implementation, we distinguish between three types: strings, numbers, and trees.   Numerics can be cast to strings, and any type can be cast to a tree.  

\sys natively enforces the rule that all $\delta$ records in $w.\Delta$ must have the same path $\pi$, because letting a widget modify arbitrary parts of a query is nonsensical and can confuse the user.   For instance, $w.\Delta=\{\diff_1, \diff_2\}$ would be rejected because their paths $\diff_1.\pi, \diff_2.\pi$ are different.

\stitle{Widget Cost Function: } The cost function $c_{WT}(w.d)$ estimates a numeric cost based on the domain $w.d$ of an instantiated widget. In general, the cost function can measure the homogeneity of the subtrees in $w.d$, the number of options, or other characteristics.  In our prototype, we assume that $c_{WT}$ is a low dimensional polynomial that increases monotonically with respect to the size of the domain $|w.d|$.  

Specifically, the cost functions used in our experiment take the form $c_{WT}(w.d) = a_0 + a_1\times|w.d| + a_2\times|w.d|^2$, where $a_i \geq 0$ and $|w.d|$ is widget domain size.    \review{Following prior interface personalization literature~\cite{gajos2007automatically},  we collected timing traces (in milliseconds) by interacting with different widget types instantiated with different domain sizes, and fit the cost function to the traces to derive the parameters $a_i$ for each widget type.  The fit model represents how efficient the widget is to use\footnote{The parameters can readily be adapted to account for widget size, or personal preferences.    For instance, if a user strongly prefers a specific widget type, its constant parameter can be set to be very low.  We anticipate that these can be learned over time by instrumenting the user's existing interfaces, or as an offline training procedure.  In future work, the cost function can be extended to support widget sizes as well~\cite{gajos2007automatically}.}.}

\begin{example}\it
  The following are fit cost functions for simple drop-down and textbox widgets:
  \begin{align*}
    c_{dropdown}(w.d) &= 276 + 125\times|w.d|^1 + 0.07 \times|w.d| ^ 2 \\
    c_{textbox}(w.d) &= 4790 
  \end{align*}
Note that the textbox cost function is a large constant because the average cost to interact with the textbox is fixed irrespective of the domain size. The cost function of drop-down is lower when the domain is small, since it is easier to directly choose from a small list than to select the textbox and type the input.  However as the domain increases, it becomes harder to find the desired option, and it is easier to simply use the textbox. 
\end{example}

\stitle{Widget Expressiveness: }We say that widget $w$ {\it expresses} \diff if their paths are the same ($w.\pi=\diff.\pi$) and the subtree $\tau_2\in w.d$ is contained in the widget's domain.   Similarly, we say a set of widgets $W$ can express a given edge $e=(q_i,q_j,t_k)$ if each $\diff\in t_k$ can be expressed by a widget in $W$.  Finally, two nodes $q_i$ and $q_j$ are connected with respect to widgets $W$ if there exists a path between them in the interaction graph such that $W$ expresses each edge in the path.

\subsection{Interactive Interfaces}  \label{s:model_interface}

An interface $I = (W_I, q_I^0)$ is a set of widgets $W_I$ and an initial query $q_I^0$, which can be any query in the interaction graph. We choose the earliest one, but can use other factors (e.g., occurrence frequency).   The user interacts with widgets to transform $q_I^0$ to other queries part of the desired analysis.  We now describe two important interface characteristics: its cost, and its expressiveness.

\stitle{Interface Cost: } There are many possible interfaces for a given query log $Q$.  For instance, we may simply create one button for every query $q_i\in Q$, where clicking on the $i^{th}$ button replaces the current query with $q_i$.  This can certainly express $Q$, but may be very undesirable if there are many queries.  For this reason, we define the cost of interface $I$ as the sum of its widget's costs\footnote{\review{In general, the cost should simply be incrementally computable.   Prior work such as Gajos et al.~\cite{gajos2007automatically} use a similar formulation.  In our case, we assume the layout is fixed, while they search across layouts as well.}}: $C_I = \sum_{w\in W} c_{w.WT}(w.\Delta)$.

\stitle{Interface Expressiveness: } 
\review{Ideally, interface expressiveness measures the ability to express the user's actual goals.  As a proxy, we define the expressiveness of interface $I$ with respect to its closure. In general, the closure $I_{closure}$ is all queries expressible by applying all possible sequences of interactions using $W_I$ to $q_I^0$.  In practice, we compute closure with respect to a log $Q$ as $I_{closure}\cap Q$, and we compute expressiveness with respect to $Q$ as $|I_{closure}\cap Q|/|Q|$.  }

\subsection{Interface Generation Problem}\label{s:problem}
We now state the main problem statement.  

\begin{problem}[Interface Generation]\label{p:prob1}
  Given a query log $Q$, a predefined library of widget types, a threshold $\gamma$ for minimum percentage of the query log to cover, generate an optimal interface $I^*$ such that:
  \begin{itemize}[topsep=-1mm, itemsep=0mm]
  \item $|I^*_{closure} \cap Q| \ge \gamma \times |Q|$
  \item $C_{I^*}$ is minimal
  \end{itemize}
\end{problem}

The problem is NP-hard, and the following is a proof sketch using reduction from vertex cover.
\begin{proof}[NP-Hardness]
  Let $G=(V,E)$ be the vertex cover graph.  We will construct an interaction graph $G' = (V', E')$ in the following way.  Each edge $e_i = (u, v)\in E$ is mapped to a vertex $s_i$ in $V'$.  We also create a dummy node $s^*$ in $V'$ and create two edges $(s_i, s^*, u)$ and $(s_i, s^*, v)$ between  each $s_i$ and $s^*$.   These edges are labeled with the incident vertex $u$ and $v$ in $V$.  Further, we define one widget type for each edge label and fix its cost function to $1$.  Thus, solving the interaction generation problem with $\gamma=1$ ensures that every vertex $v'\in V'$ is expressed in the interface closure (e.g., each edge in $E$ is covered), and the set of widgets that are selected has minimal cost, which minimizes the number of vertices selected in the vertex cover problem.
\end{proof}

\stitle{Discussion: }
Note that our definition of widgets simply specifies a path and domain, and is not bound any specific visual representation.  This allows \sys to be easily extended to new interaction components or even different modalities such as voice or touch gestures~\cite{Zhang2018PrecisionIF}.

Since \sys operates at the syntactic level, certain combinations of AST transformations might lead to non-executable queries.  Although this is unlikely for common transformations such as adding expression clauses or tuning parameters, it is still possible.  One solution is to speculatively parse and execute queries in the interface's closure, and visually disallow interactions that lead to these ASTs.  If the space of queries is small, this can be a way to both verify and pre-compute results for performance purposes.    

In this paper, we set $\gamma=1$, so that the entire query log can be expressed.  We also find that a given interface can often generalize to express queries not present in the log.  There are two reasons for this: 1) widgets such as sliders can express more than the subtrees that they are initialized with, and 2) transformations expressible by any combination of widgets are possible, thus the number of expressible queries potentially increases combinatorially with the number of widgets in the interface.  Our experiments will evaluate this generalization capability for different query logs.

\section{Interaction Mapper}\label{s:interface}

Due to the NP-hardness of the interface generation problem, we now describe a heuristic solution based on graph contraction.    The heuristic is split into two phases. Initialization constructs an initial interface that can express all queries in the log.  However, it likely has high cost and contains redundant widgets that express overlapping sets of edges in the interaction graph.    Thus, the merging phase greedily merges and removes redundant widgets to simplify the interface and reduce the cost.

\begin{algorithm}[t]\small
	\begin{algorithmic}[1]
		\State I = $\emptyset$
		\For{each $\diff$ in $\Omega$}
        \State $\Omega_{\diff.\pi}$.add(\diff)
		\EndFor
		\For{each $\Omega_\pi$}
        \State I.add(pickWidget($\Omega_\pi$, $L$))
		\EndFor
		\State \Return I
	\end{algorithmic}
    \caption{Initialize($\Omega$, $L$)} 
    \label{a:init}
\end{algorithm}

\begin{algorithm}[t]\small
\begin{algorithmic}[1]
  \State $d = \cup_{\delta\in \Omega_\pi} \{\delta.\tau_1, \delta.\tau_2\}$\hfill // get subtrees
  \State $L' = \{WT\in L | WT.r(d)\}$ \hfill // valid widget types
  \State $WT^* = \argmin_{WT\in L'} WT.c(d)$\hfill // lowest cost widget type
  \State \Return $WT^*(\pi, d)$
\end{algorithmic}
\caption{pickWidget($\Omega$, $L$)} 
\label{a:pick}
\end{algorithm}

\subsection{Initialization}	

To create the initial interface, Algorithm~\ref{a:init} naively clusters edges in the interaction graph and selects a widget type to initialize for each cluster.   Let $\Omega = \difftable$ be the set of {\diff}s in the interaction graph.  Let us then partition $\Omega$ based on each \diff's path $\pi$, thus partition $\Omega_\pi = \{ \diff\in\Omega | \diff.\pi = \pi \}$ contains the {\diff}s with the same path.      Although we could use finer-grained partitions (e.g., one partition per edge),  we find that our approach improves the speed considerably and often results in comparable interfaces in practice.

We then instantiate a widget for each partition by calling \texttt{pickWidget($\Omega_\pi$, $L$)}, passing in the partition along with a library of widget types $L$.  This function first extracts the domain $d$ (set of subtrees) defined by the partition, and checks whether each widget type's rule accepts $d$.   Among these, it instantiates the lowest-cost widget type $WT^*$.  This guarantees that there is at least one widget for every edge, and that every query is within the interface's closure.

\begin{figure}[h]
  \centering
  \includegraphics[width=\columnwidth]{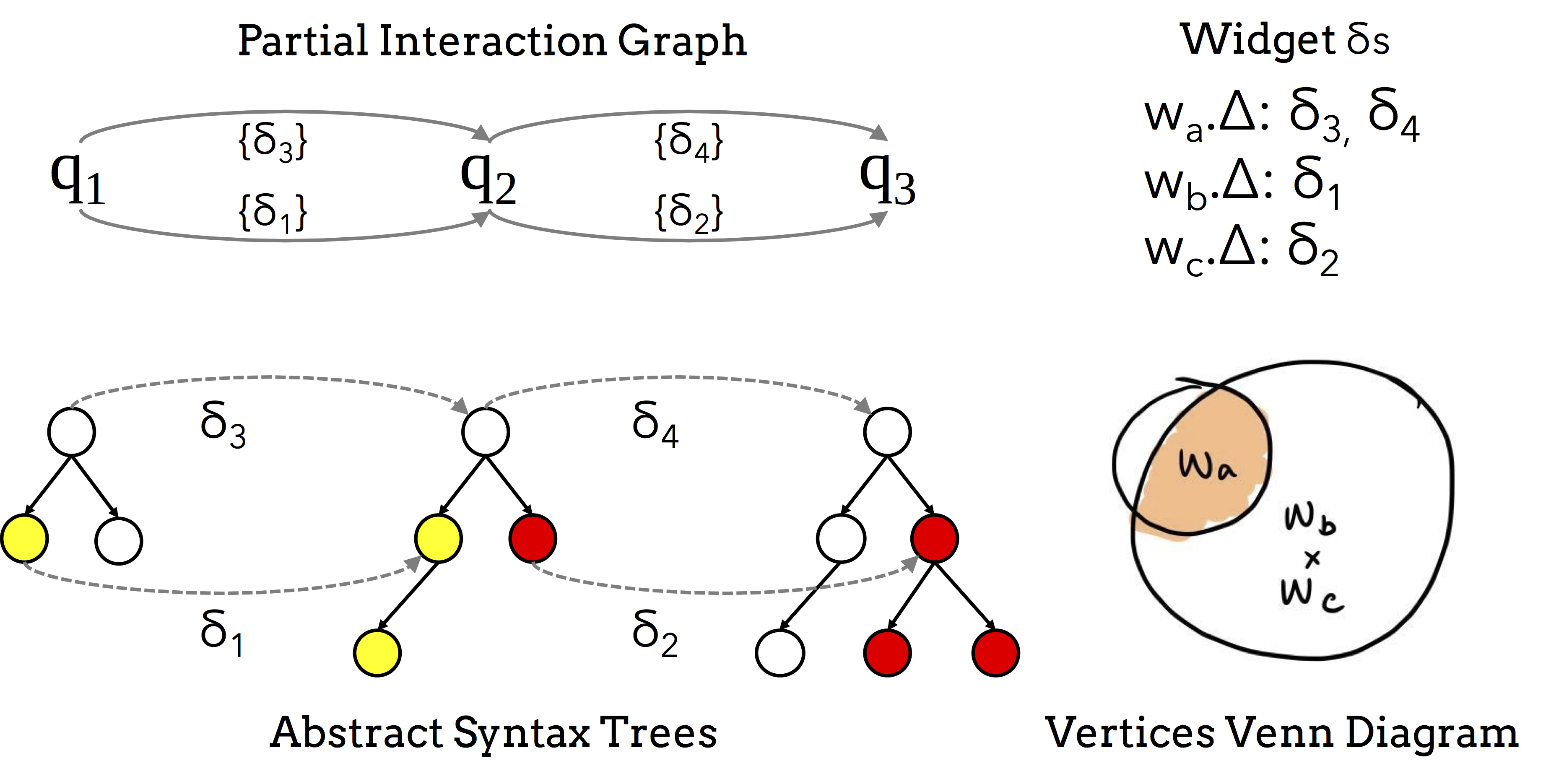}
  \caption{\small Differences between ($q_1$, $q_2$), and ($q_2$, $q_3$). The interaction graph (top);  the ASTs and {\diff}s (bottom).}
  \label{f:prunetree}
\end{figure}

\begin{example}\it
  Figure~\ref{f:prunetree} depicts part of the interaction graph (top) and the AST differences (bottom) between three queries. $q_1\rightarrow q_2$ differ in the yellow subtree $\diff_1$, which appears as the edge $\{\diff_1\}$ in the interaction graph.  Its ancestor $\diff_3$ is also included in the graph because replacing the entire AST is a viable transformation.    

  This graph is initialized with three widgets.  $w_a$ is initialized with $\{\diff_3,\diff_4\}$, and its domain $w_a.d$ consists of the three ASTs.  Thus, the user can interact with $w_a$ to select one of the three ASTs to replace the root node of the interface's current query.   $w_b$ is initialized with $\{\diff_1\}$ and its domain consists of the two yellow subtrees,   while $w_c$ is initialized with $\{\diff_2\}$ and its domain contains the two red subtrees.
\end{example}

\noindent Clearly, not all three widgets are needed to express the three queries: $w_a$ alone is sufficient but can only express the three queries, whereas the pair $(w_b,w_c)$ is sufficient but can express any combination of the yellow and red subtrees.  To this end, we use a merging procedure to reduce this redundancy.

\begin{algorithm}[t]\small
  \begin{algorithmic}[1]
    \State // Compute vertices incident to widgets' edges 
    \State $V_a = \cup_{\diff\in w_a.\Delta} \{\diff.q_1, \diff.q_2\}$ 
    \State $V_d = \cup_{w_d\in W_d, \diff\in w_d.\Delta} \{\diff.q_1, \diff.q_2\}$ 
    \State $V = V_a \cap V_d$ \\

    \State // Get {\diff}s where both incident queries are in intersection $V$
    \State $g_a = \{ \diff\in w_a.\Delta | \diff.q_1\in V \wedge \diff.q_2\in V \}$
    \State $g_d = \cup_{w_d\in W_d}\{ \diff\in w_d.\Delta | \diff.q_1\in V \wedge \diff.q_2\in V \}$ \\

    \State // Cost reductions if {\diff}s removed from descendents
    \State $s_d = 0$
    \For{each $w_d$ in $W_d$}
      \State $w' = pickWidget(w_d.d - g_d)$ 
      \State $s_d = s_d + w_d.cost - w'.cost$   
    \EndFor

    \State // Cost reductions if {\diff}s removed from ancestor
    \State $w' = pickWidget(w_a.d - g_a)$
    \State $s_a = w_a.cost - w'.cost$\\

    \If{$s_a > s_d$}  
      \State // Remove overlapping {\diff}s from ancestor
      \State \Return (pickWidget($w_a.d - g_a$), $W_d$)
    \Else
      \State // Remove overlapping {\diff}s from descendents
      \State $W' = \{ pickWidget(w_d.d - g_d) | w_d\in W_d  \}$
      \State \Return ($w_a$, $W'$)
    \EndIf
  \end{algorithmic}
  \caption{Merge($w_a$, $W_d$)} \label{a:merge}
\end{algorithm}

\subsection{Merging}

Merging removes widget redundancy while ensuring that every query can still be expressed. This is described in Algorithm~\ref{a:merge}, which iteratively compares pairs of widgets $w_i$ and $w_j$ for which $w_i.\pi$ is a prefix of $w_j.\pi$, and merges them.  In Figure~\ref{f:prunetree} the path for $w_a$ is a prefix of $w_b$'s path. We do not consider comparing other pairs of widgets (e.g., $w_b$ and $w_c$) because their paths would refer to non-overlapping parts of the query AST, and would not make sense to merge together.  Note that widgets with this prefix relationship are very common in the interaction graph, because every ancestor of a subtree is logically added as a transformation in \difftable.    

At a high level, the algorithm compares a widget and a set of descendent widgets (e.g., $w_a$ vs $(w_b, w_c)$). Since they are redundant, they express edges that connect the same pairs of vertices in the graph.  This is depicted as the venn diagram---the overlapping edges (colored in orange) will be exclusively assigned to the ancestor or descendent widgets based on the resulting interface's cost.

Algorithm~\ref{a:merge} chooses between an ancestor widget $w_a$ and the set of descendents $W_d$ (e.g., $w_b, w_c$).    Lines 2-4 identify the vertices incident to the edges that each widget expresses.  Recall that each edge is a set of {\diff}s, thus if a widget expresses any \diff in the edge, the incident vertices are counted.  The intersection $V$ is thus the vertices expressed by both widget options.  Lines 7-8 then identify the intersecting {\diff}s whose vertices are both within $V$---these are the candidates to exclusively assign to the ancestor or descendent widgets.  Lines 11-17 compute the cost change to remove the intersecting {\diff}s from the widgets; lines 19-25 pick the option that most reduces the cost.

We iteratively perform this merging procedure until the resulting interface cost does not reduce anymore.

\subsection{Generating Interfaces}\label{s:geninterface}

After generating $I^*$, an editor interface renders the widgets in a grid.  The user can optionally edit, add labels, or change the widget type for each widget.   The editor lets users modify the layout and sizes of the widgets, or a standard layout algorithm could be run~\cite{sears1993layout}.   We then compile the interface into a web application that executes an internal query $q$ by running the provided \texttt{exec()} function, and renders the results using the user provided \texttt{render()} method.  When the user interacts with widget $w$, the widget state corresponds to a value or subtree in the widget's domain.  That value is swapped into the current query at the path $w.\pi$.

\section{Optimization}\label{s:opt}

Our baseline implementation takes as input a list of XML parse trees created by a third-party query parsing service\footnote{\url{http://www.sqlparser.com/}}.  It then computes tree alignments between all pairs of ASTs, extracts all subtree differences and their ancestors, and adds them to the interaction graph.  After that, it simply executes the interaction mapper heuristic.  Our experiments show that the primary costs are in performing pairwise tree comparisons and the iterative widget merging procedure, both due to the large size of the interaction graph.  We developed two effective optimizations that reduce both overheads.

\subsection{Sliding Window Analysis}

In practice, query logs contain an ordered list of queries and often meta-data such as the session, user, and timestamp.   If the queries were generated as part of an analysis, then it is reasonable to assume that queries exhibit locality.  For instance, a user is more likely to want to care about changes between queries near each other in the log, as opposed to queries separated by 100s of other queries.

We thus pass a sliding window of size $n_{win}$ over the input query log, and only extract structural differences between pairs of queries within the window.    This optimization both reduces the number of comparisons that we need to make from $O(|Q|^2)$ to $O(|Q|*n_{win})$, and reduces the size of the interaction graph that the mapper needs to process.

\subsection{Pruning Tree Differences}

Section~\ref{s:interface} reduces the interface cost by iteratively merging redundant widgets that express overlapping subsets of the interaction graph.  In some cases, it is possible to directly prune subtree transformations when generating the interaction graph if it won't affect the resulting interface.  
  
We consider the subtree differences found by the tree alignment algorithm as leaf-{\diff}s.   Least Common Ancestor (LCA) Pruning removes all {\diff}s that are neither a leaf-\diff, nor least common ancestor of two leaf-{\diff}s.     The intuition is that a leaf-\diff  can potentially be a literal type, and be mapped to lower cost widgets such as a slider.  In contrast, its ancestors are \texttt{tree}-types with strictly higher cost; this is only justified if the ancestor can express more transformations than the leaf-\diff alone.  This only occurs at subtrees rooted at the least common ancestor of pairs of leaf-{\diff}s.

\begin{example}\it
In Figure~\ref{f:asts}, \texttt{StrExpr} is a leaf-\diff which is a literal type, whereas its ancestor \texttt{BiExpr} node is a tree type.  For the same domain size, a widget for a literal type will be the same or lower cost than a widget for a tree type.  Thus, the mapping algorithm will never select the \texttt{BiExpr} transformation.  In contrast, we need to consider the root \texttt{SELECT} difference because it also expresses the \texttt{ColExpr} transformation.  
\end{example}

\section{Experiments}\label{s:exp}

We seek to understand the cost trade-offs when generating interfaces, how well the interfaces can express queries from similar or different analyses, the effects of different query log compositions, and system runtime.   Our user study compares \sys with the original SDSS search form (``SDSS interface''); we  find that \sys creates new widgets for analyses that are challenging to express in the SDSS interface, and is initially easier to understand.  Experimental details can be found in \Cref{a:exp_recall_crossclient,a:exp_perf,a:exp_user,a:exp_precision}.

\stitle{Query Logs:} We used 3 SQL query logs that differ in the variety and regularity of changes between queries, and describe each below.  Our sample of the \textbf{Sloan Digital Sky Survey (SDSS)}~\cite{sdss} query log~\cite{sdsslog} contains $127,461$ queries submitted to the SDSS sky server database between 11/27/2004 and 11/30/2004, along each query's client IP ($286$ unique clients). We partition the queries by client, and  assume each client represents one analysis session.  Although some clients have more than $10,000$ queries, most are far fewer.  Our experiments use random clients containing $\ge200$ queries in their log.  \Cref{q:sdss} shows a sample of queries from a single user; the queries for each user are considerably different, but the {\it changes} between a given user's queries are very similar and highly structured.

{\small
\begin{lstlisting}[
	caption={Sample of SDSS queries},
	label={q:sdss}
]
    SELECT * FROM ^\highlight{SpecLineIndex}^ WHERE specObjId=^\highlight{0x400}^;
    SELECT * FROM ^\highlight{XCRedshift}^ WHERE specObjId=^\highlight{0x199}^;
    SELECT * FROM ^\highlight{SpecLineIndex}^ WHERE specObjId=^\highlight{0x3}^;
\end{lstlisting}
}

The \textbf{OLAP} synthetic log contains 200 queries generated by a random walk through the OLAP query space; each step adds, removes, or modifies a random dimension, aggregation, or filter.  \Cref{q:olap} shows 3 example queries: an aggregation is removed, then a predicate is changed.  The type of transformation is more rich than in the SDSS logs, but simpler than those for the ad-hoc log below.

{\small
	\begin{lstlisting}[
	caption={Synthetic OLAP queries},
	label={q:olap}
	]
    q1 = SELECT ^\highlight{COUNT(Delay)}^, DestState FROM ontime 
          WHERE Month=9 and Day=3
       GROUP BY DestState;
    q2 = SELECT DestState FROM ontime 
          WHERE Month=^\highlight{9}^ and Day=3
       GROUP BY DestState;
    q3 = SELECT DestState FROM ontime 
          WHERE Month=^\highlight{8}^ and Day=3
       GROUP BY DestState;
	\end{lstlisting}
}

The \textbf{Ad-hoc} query log contains queries generated by students during open-ended exploration using Tableau of the OnTime flight delays dataset~\cite{ontime}.  We treat each student's queries as a separate log.  There is considerable variation in queries and changes in this log. 

\stitle{Implementation} We implemented parsing and interaction mining in Java, widget mapping and rendering in Python, and generated interfaces in HTML+JavaScript.  We defined 9 HTML widget types natively supported in modern browsers: text-box, toggle-button, single checkbox, radio button, drop-down list, slider, range slider, checkbox list, drag-and-drop.  Their cost functions are learned as described in~\Cref{s:model_widgets}.   We manually created \texttt{exec()} and \texttt{render()} functions; we labeled and repositioned the generated widgets for presentation purposes. Experiments are run on a MacBook Air with Intel Core i5 1.6 GHz CPU and 8GB RAM.  

{\small
\begin{lstlisting}[
	caption={Sample of ad hoc student queries.},
	label={q:adhoc}
]
  q1 = SELECT CAST(uniquecarrier) AS uniquecarrier 
         FROM ontime;
  q2 = SELECT SUM(flights) FROM ontime 
        WHERE canceled = 1 
       HAVING SUM(lights) > 149 and SUM(flights) < 1354;
  q3 = SELECT (CASE carrier 
               WHEN 'AA' THEN 'AA' 
               ELSE 'Other' END) AS carrier,
              FLOOR(distance/5) AS distance
         FROM ontime;
\end{lstlisting}
}

\subsection{Interface Mapping Tradeoffs}
\label{sec:case_stud}

\begin{figure*}[h]
  \centering
  \begin{subfigure}[t]{0.19\textwidth}
    \centering
    \includegraphics[width=.9\textwidth]{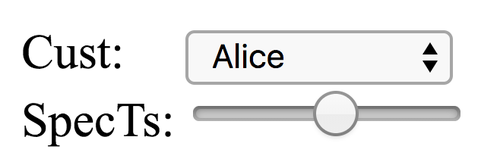}
    \caption{\Cref{q:qlog1}}
    \label{fig:qlog1}
  \end{subfigure}
  \hfill
  \begin{subfigure}[t]{0.19\textwidth}
    \centering
    \includegraphics[width=.5\textwidth]{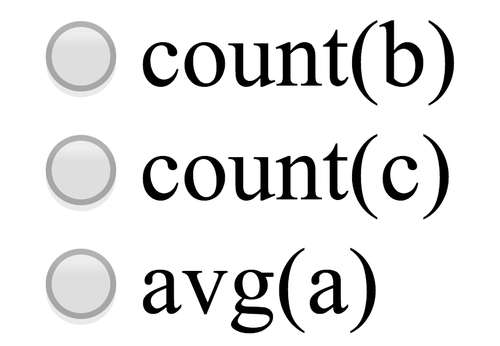}
    \caption{\Cref{q:qlog2} (left)}
    \label{fig:qlog2_less}
  \end{subfigure}
  \hfill
  \begin{subfigure}[t]{0.19\textwidth}
    \centering
    \includegraphics[width=.8\textwidth]{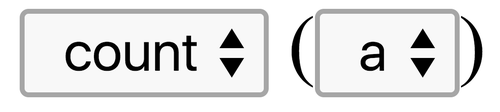}
    \caption{\Cref{q:qlog2} (right)}
    \label{fig:qlog2_more}
  \end{subfigure}
  \hfill
  \begin{subfigure}[t]{0.19\textwidth}
  \includegraphics[width=\textwidth]{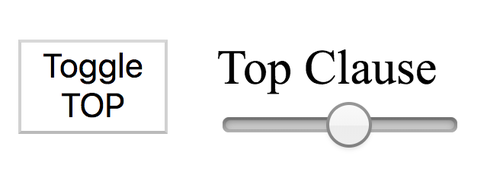}
  \caption{\Cref{q:qlog3}}
  \label{fig:qlog3}
  \end{subfigure}
  \hfill
  \begin{subfigure}[t]{0.19\textwidth}
	\centering
	\includegraphics[width=\textwidth]{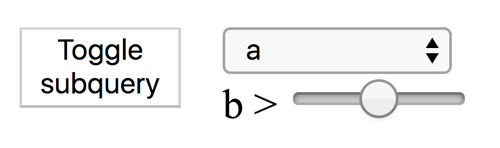}
    \caption{\Cref{q:qlog4}}.
	\label{fig:qlog4}
  \end{subfigure}
  \caption{Widgets mapped to different example logs to illustrate:  (a) ancestor-descendent trade-off made during Merging, (b) single widget from low variety log,  (c) multiple widgets from higher variety log, (d) adding a \texttt{TOP} clause and updating its limit, (e) adding a subquery and modifying it. }
\end{figure*}

\sys is able to generate interfaces for simple query changes, such as modifying a single numeric threshold in a predicate, or complex query changes, such as adding a subquery and then modifying parts of that subquery.    In this subsection, we showcase different trade-offs that \sys makes depending on the composition and types of queries in simple synthetic query logs.  The purpose is to show that the interfaces that \sys generates are based on the complexity of the query {\it changes}, and not the underlying query complexity.  For space constraints, we will only show the interaction widgets in the interface since the output visualization is the same.

{\small
\begin{lstlisting}[
  caption={Simple parameter changes (\highlight{highlighted}) to a complex query. },
  label={q:qlog1}
]
    SELECT spec_ts,sum(price) FROM (
       SELECT action,sum(customer) FROM t 
       WHERE spec_ts > now and spec_ts < now + ^\highlight{3}^
     )
     WHERE cust = ^\highlight{'Alice'}^ and country = 'China' 
    GROUP BY spec_ts;
\end{lstlisting}
}  
\subsubsection{Simple Parameter Changes}
We first showcase simple parameter changes in complex queries.  \Cref{q:qlog1} shows the query template we used to create an example query log.  The template contains a subquery and multiple predicates, and we modified the literal \highlight{3} in the subquery's \texttt{spec\_ts} predicate, and the customer name from  \highlight{'Alice'} to other names such as 'Bob'.

Figure~\ref{fig:qlog1} shows the widgets generated from this query log.  It contains a single widget for each of the two types of changes: a drop-down list to select the small number of customer names, and a slider to vary the range size in the \texttt{spec\_ts} predicate.    Note that we could have mapped a drop-down or other widget to  the numeric changes, but a slider matches the numeric type and has a lower cost.  Further, other possible changes to the query, such as changing the SELECT clause, adding to the FROM clause, changing the \texttt{country} name are not present in the query log, and thus are not needed to be expressed in the interface. 

Also note that this interface can express queries not present in the query log.  For instance, the combination of \texttt{cust = 'Bob'} and \texttt{spec\_ts < now + 9} was not present in the log, but can be expressed.  In fact, the interface can express the cross-product of the widgets' domains.

\subsubsection{Adaptivity to Query Log Size }
We now use a trivial query structure, shown in \Cref{q:qlog2}, to show how the composition of the query log changes the widgets that are selected.   We use a single function call, and simply change the function name and its single argument.  The left side shows three queries used to generate the interface in \Cref{fig:qlog2_less}, whereas we then appended the queries on the right side to generate the interface in \Cref{fig:qlog2_more}

{\small
\begin{lstlisting}[
  caption={Function name and argument varies.},
  label={q:qlog2},
]
   -- Qs for ^Figure~\ref{fig:qlog2_less}^      -- Qs added for ^Figure~\ref{fig:qlog2_more}^
   SELECT ^\highlight{avg}^(^\highlight{a}^)         SELECT ^\highlight{avg}^(^\highlight{b}^) 
   SELECT ^\highlight{count}^(^\highlight{b}^)        SELECT ^\highlight{count}^(^\highlight{a}^)
   SELECT ^\highlight{count}^(^\highlight{c}^)        SELECT ^\highlight{avg}^(^\highlight{c}^)
                           SELECT ^\highlight{avg}^(^\highlight{d}^)
                           SELECT ^\highlight{avg}^(^\highlight{e}^)
                           SELECT ^\highlight{count}^(^\highlight{d}^)
                           SELECT ^\highlight{count}^(^\highlight{e}^)
                           SELECT ^\highlight{count}^(^\highlight{b}^)
                           SELECT ^\highlight{count}^(^\highlight{c}^)
                           SELECT ^\highlight{avg}^(^\highlight{a}^)
\end{lstlisting}
}

\noindent With three queries, it is easier to directly choose the query that the user is interested in, and \sys maps the queries to a radio box selection widget (\Cref{fig:qlog2_less}).  The domain of this widget contains the full AST trees for each query, and selecting an option simply replaces the current query with the selected AST.  However when there are more queries using a single widget becomes unwieldy---it is difficult to choose from a long list of 10 options.  In this case, \Cref{fig:qlog2_more} creates a separate widget for each component of the function that changes, which reduces the number of options in each drop-down at the cost of adding a second widget.

{\small
\begin{lstlisting}[
  caption={Queries first add TOP clause, then modify it.},
  label={q:qlog3},
]
q1 = SELECT g.objID 
       FROM Galaxy as g, 
            dbo.fGetNearbyObjEq(5.848,0.352,2.0616) as d 
       WHERE d.objID = g.objID;
q2 = SELECT ^\highlight{TOP 1}^ g.objID 
       FROM Galaxy as g, 
            dbo.fGetNearbyObjEq(5.848,0.352,2.0616) as d 
       WHERE d.objID = g.objID; 
q3 = SELECT TOP ^\highlight{10}^ g.objID 
       FROM Galaxy as g, 
            dbo.fGetNearbyObjEq(5.848,0.352,2.0616) as d 
       WHERE d.objID = g.objID;
\end{lstlisting}
}

\subsubsection{Structural differences}

\Cref{q:qlog3} shows SDSS queries where a \texttt{TOP} clause is first added, and then the limit is modified. The query is quite complex and contains UDFs and multiple tables.  The changes are much simpler, albeit more complex than simple parameter changes. \Cref{fig:qlog3} shows that \sys generates a single \texttt{Toggle TOP} button to toggle the presence of the \texttt{TOP} clause, and then a slider to select the number of records to return;  the slider is only active when the \texttt{TOP} clause is enabled.    Note that \sys does not understand the semantics of the top clause, and identifies these changes syntactically.

As a final example, \Cref{q:qlog4} shows an example where a subquery is added to the \texttt{FROM} clause, and parts of the subquery are subsequently modified.  
\Cref{fig:qlog4} shows the mapped widgets. A button toggles between the table \texttt{T} and the subquery.  When the subquery is toggled, the drop-down list and slider are enabled to modify the projection and predicates in the subquery.  

{\small
	\begin{lstlisting}[
	caption={Adding, then modifying a subquery.},
	label={q:qlog4},
	]
  q1 = SELECT * FROM ^\highlight{T}^;
  q2 = SELECT * FROM (SELECT ^\highlight{a}^ FROM T WHERE b > ^\highlight{10}^);
  q3 = SELECT * FROM (SELECT ^\highlight{a}^ FROM T WHERE b > ^\highlight{20}^);
  q4 = SELECT * FROM (SELECT ^\highlight{b}^ FROM T WHERE b > ^\highlight{20}^); 
	\end{lstlisting}
}

{\it \stitle{Takeaways: } \sys generates widgets independent of the underlying query complexity.  It supports a range of transformations beyond simple parameter changes, including subquery transformations.  It also makes trade-offs between widget complexity and the number of widgets. }

\begin{figure*}[h]
  \centering
  \begin{subfigure}[t]{0.29\textwidth}
	\centering
	\includegraphics[width=\textwidth]{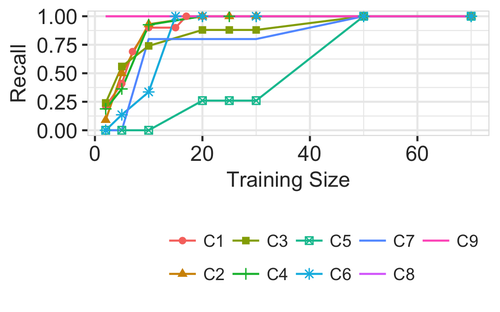}
	\caption{Recall: SDSS log}
	\label{fig:recall}
  \end{subfigure}\hfill
  \begin{subfigure}[t]{0.20\textwidth}
	\centering
	\includegraphics[width=.9\textwidth]{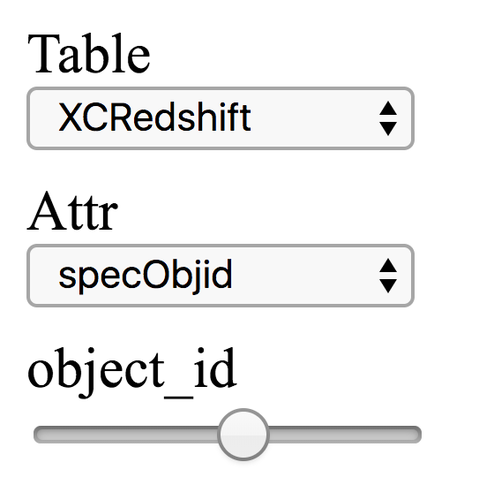}
	\caption{Widgets: SDSS user \texttt{C1}}
	\label{fig:sdss_interface}
  \end{subfigure}\hfill
  \begin{subfigure}[t]{0.29\textwidth}
	\centering
	\includegraphics[width=\textwidth]{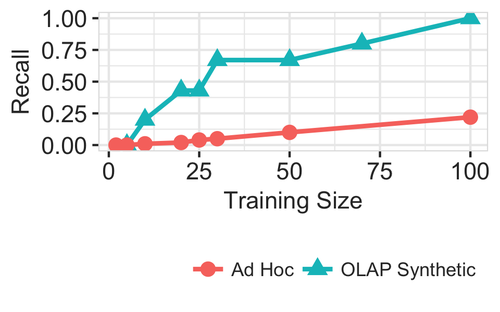}
	\caption{Recall: OLAP and ad-hoc logs}
	\label{fig:recall_both}
  \end{subfigure}\hfill
  \begin{subfigure}[t]{0.18\textwidth}
	\centering
	\includegraphics[width=.75\textwidth]{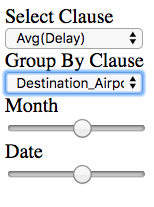}
	\caption{Widgets: OLAP log }
	\label{fig:synthetic_interface}
  \end{subfigure}
  \caption{(a) Recall for SDSS client logs. (b) Interface generated for client \texttt{C1}. (c) Recall for synthetic OLAP and ad-hoc student exploration logs. (d) Interface generated for synthetic query log.}
\end{figure*}

\subsection{Interface Generalizability}\label{s:exp_recall}

\Cref{s:model_interface} defined a strict definition of expressiveness as the percentage of the query log that the interface can express (is within its closure).  We extend this to measure how well a generate interface can express future (unseen) queries from the same analysis.  For an input log of size $n$, we split it into $n_{holdout}$ ``hold-out'' queries and $n_{training}$ ``training'' queries.  We run \sys over a subset of the training queries, and compute the fraction of the hold-outs that the generated interface can express. This is called {\it recall} in machine learning.  If the recall is high, it suggests that other queries in the log's analysis are expressible in the generated interface.  We report the {\it rate} that the recall reaches 100\% as the subset of the training set increases, and vary the composition of the input log as described below.

\subsubsection{Single-Client SDSS Logs}
In this experiment, we use 9 random SDSS client logs (the others are similar), and evaluate each one in isolation.  Since the logs have varying length, we partition each log into 200-query windows, use the first $n_{training}\in[1,100]$ queries as training, and the last $100$ as hold-out.  We report the average recall over the windows. 

\Cref{fig:recall} shows the recall for each client log as the training size varies (x-axis).    10 queries is sufficient to express the hold-out queries for the majority of client logs,  and 50 training queries increases recall to 100\%.   There is one user---C5---for which the recall increases slowly over the first 50 training queries.  The reason is because the query structure does not change very much, however some of the literal values that the user changes are never encountered in the first 50 queries.  We expect that extending \sys to leverage the database schema and contents can drastically improve the recall curves, and leave this as future work.

\Cref{q:sdss} shows a sample of SDSS queries from client C1. C1 looks up information about objects from tables containing spectral line or red shift data.  Both tables use the same attribute name to index the objects, so the user primarily changes the table and specifies the ID.  \sys identifies that the ID is numeric, and generates a simple slider (\Cref{fig:sdss_interface}).  Not shown in the example queries is that the field name may switch to a different ID attribute.  Thus the interface creates widgets to change the table, attribute name, and ID.    We see how a given analysis for a given user can be simple, even though the database is complex.

\subsubsection{Single-Client OLAP and Ad-hoc Logs}
\Cref{fig:recall_both} plots the recall curve for the synthetic OLAP log (blue) and the average over the student logs (red).  The reason the OLAP curve increases slower than in the SDSS dataset is because many different parts of the query---the grouping, aggregation, and predicates---may change within the same analysis log.  In contrast, the SDSS client's analysis is localized and more repetitive.  Thus it requires more training queries to predict the latter 100 queries in the log.  Figure~\ref{fig:synthetic_interface} shows the selected widgets using the first 100 queries as input. Two drop-downs express the ways that the aggregation and grouping clauses changed in the log, while sliders express the predicate modifications.  Note that only 50 queries were needed to map the same widgets, however more queries were needed to fill the widgets' domains (e.g., drop-down options).

Interfaces are not guaranteed to generalize.  The red line in \Cref{fig:recall_both} shows that the recall is quite low: even with 100 training queries, the interface only expresses $\approx20\%$ of the 100 hold-out queries.  \review{\sys is suited for analyses that involve a closed set of query transformations, and may not be suitable for ad-hoc, non-repetitive settings.}

\begin{figure*}
  \centering
  \begin{subfigure}[t]{0.32\textwidth}
	\centering
	\includegraphics[width=0.9\textwidth]{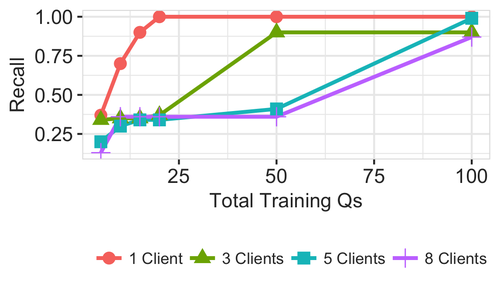}
	\caption{}
	\label{fig:sdss_multiclient_total}
  \end{subfigure}\hfill
  \begin{subfigure}[t]{0.30\textwidth}
	\centering
	\includegraphics[width=0.9\textwidth]{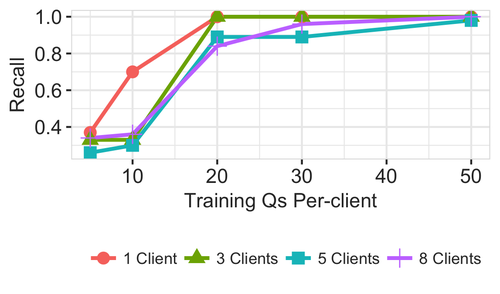}
    \caption{}
	\label{fig:sdss_multiclient_perclient}
  \end{subfigure}\hfill
  \begin{subfigure}[t]{0.30\textwidth}
	\centering
	\includegraphics[width=0.9\textwidth]{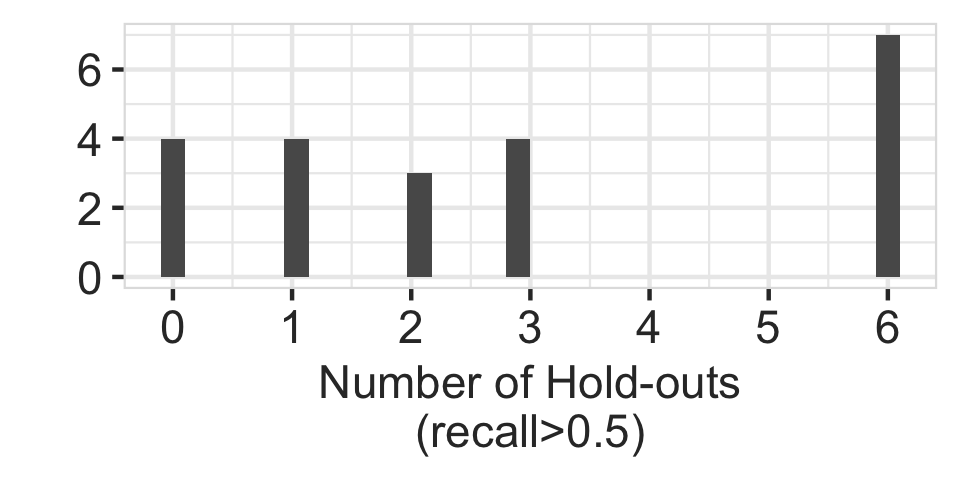}
    \caption{}
	\label{fig:sdss_crossclient_dotplot}
  \end{subfigure}\hfill
  \vspace*{-.1in}
  \caption{(a) Multi-client SDSS. Vary total training queries.  (b) Multi-client SDSS.  Vary training queries per-client. (c) Cross-client SDSS.}
\end{figure*}

\subsubsection{\review{Multi-Client SDSS Logs}}\label{s:exp_recall_multiclient}
\review{We now study recall under heterogeneous conditions.  We selected and interleaved $M\in\{1, 3, 5, 8\}$ random client logs (truncated to $200$ queries each).   We then pick $50$ queries as hold-out, and vary the size of the training data in two ways.  \Cref{fig:sdss_multiclient_total} varies the {\it total} number of training queries from 5 to 100 (x-axis). The recall increases very slowly because there are simply fewer examples from each client, similar to the ad-hoc logs above with high query variability.   Second, \Cref{fig:sdss_multiclient_perclient} varies the number of training queries {\it per client} ($10$ means $10M$ total training queries).  In this case, we see that recall increases rapidly, similar to the single-client experiments, since each client is still quite simple.  } 

\subsubsection{\review{Cross-Client SDSS Logs}}\label{s:exp_recall_crossclient}
\review{A precision interface is specialized to a specific analysis, and we now study the extent that an interface from one client's logs $Q_i$ are able to express queries from other client logs $Q_j, j\ne i$.  We expect near-zero recall for most clients, and high recall for clients that perform similar analyses.  To do so, we used the $M=22$ SDSS client logs that contain $\ge100$ or more queries, truncated them to $100$ queries each, used $Q_i$ to generate an interface, and measured its recall for each of the other logs.   }

\review{\Cref{fig:sdss_crossclient_dotplot} summarizes the results as a histogram.  The bar at x-axis value of 1 means that for 4 training client logs (y-axis), the interface had $recall>0.5$ for 1 hold-out logs (excluding the training log).  We see that the majority of training clients generated interfaces that benefit at least one other client.  In fact, 7 interfaces were able to express 6 other clients.   \Cref{a:exp_recall_crossclient} plots the pair-wise recall matrix for all 22 clients as a heat map. }

{\it \stitle{Takeaways: } \sys successfully identifies systematic structural transformations in query logs and generates simple, precise interfaces that can express future analysis queries with a few dozen training examples. Its interfaces do not generalize if the query variation is ad-hoc.  }

\subsection{Runtime}
\label{sec:atscale}

For space constraints, we summarize the runtime experiments reported in \Cref{a:exp_perf}.  We evaluated the runtime and optimizations using the SDSS query log.  The sliding window and LCA pruning optimizations help reduce the runtime by multiple orders of magnitude without changing the output interfaces.  \sys scales to 10,000 queries and runs within 10 seconds. On logs of $\approx 2000$ queries, \sys runs within 3 seconds.

\begin{figure*}[t!]
    \centering
  \begin{subfigure}[b]{0.28\textwidth}
    \centering
    \includegraphics[width=0.9\textwidth]{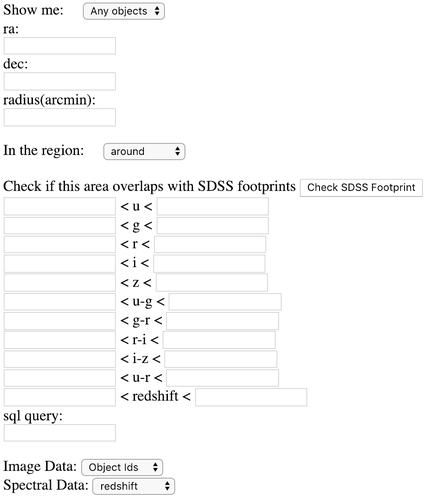}
    \caption{SDSS interface with styling and output controls removed.}
    \label{fig:ui-1}
  \end{subfigure}\hfill
  \begin{subfigure}[b]{0.18\textwidth}
	    \includegraphics[width=\textwidth]{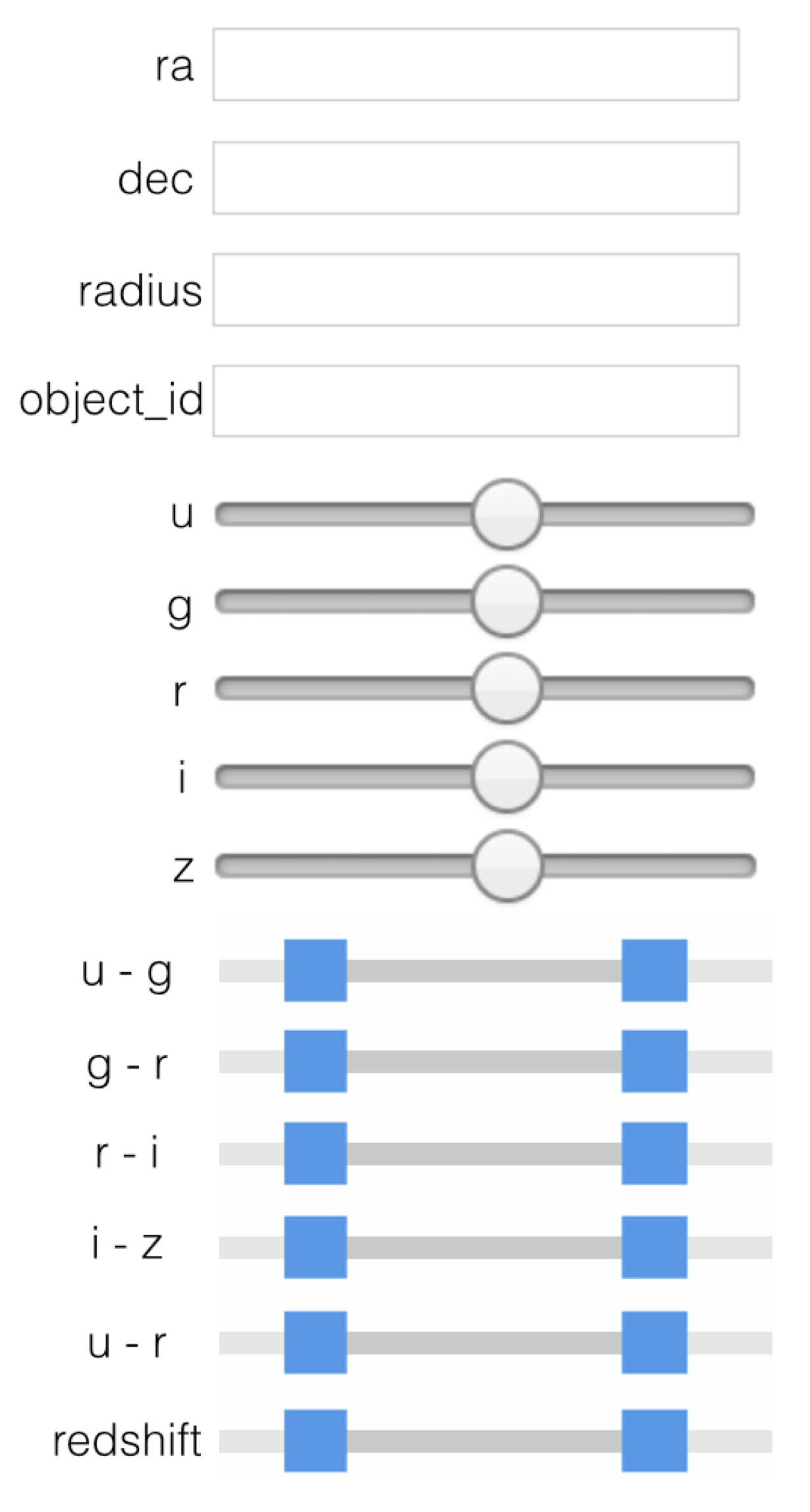}
	    \caption{Precision Interface}
	    \label{fig:ui-2}
  \end{subfigure}\hfill
  \begin{subfigure}[b]{0.4\textwidth}
    \centering
    \includegraphics[width=\textwidth]{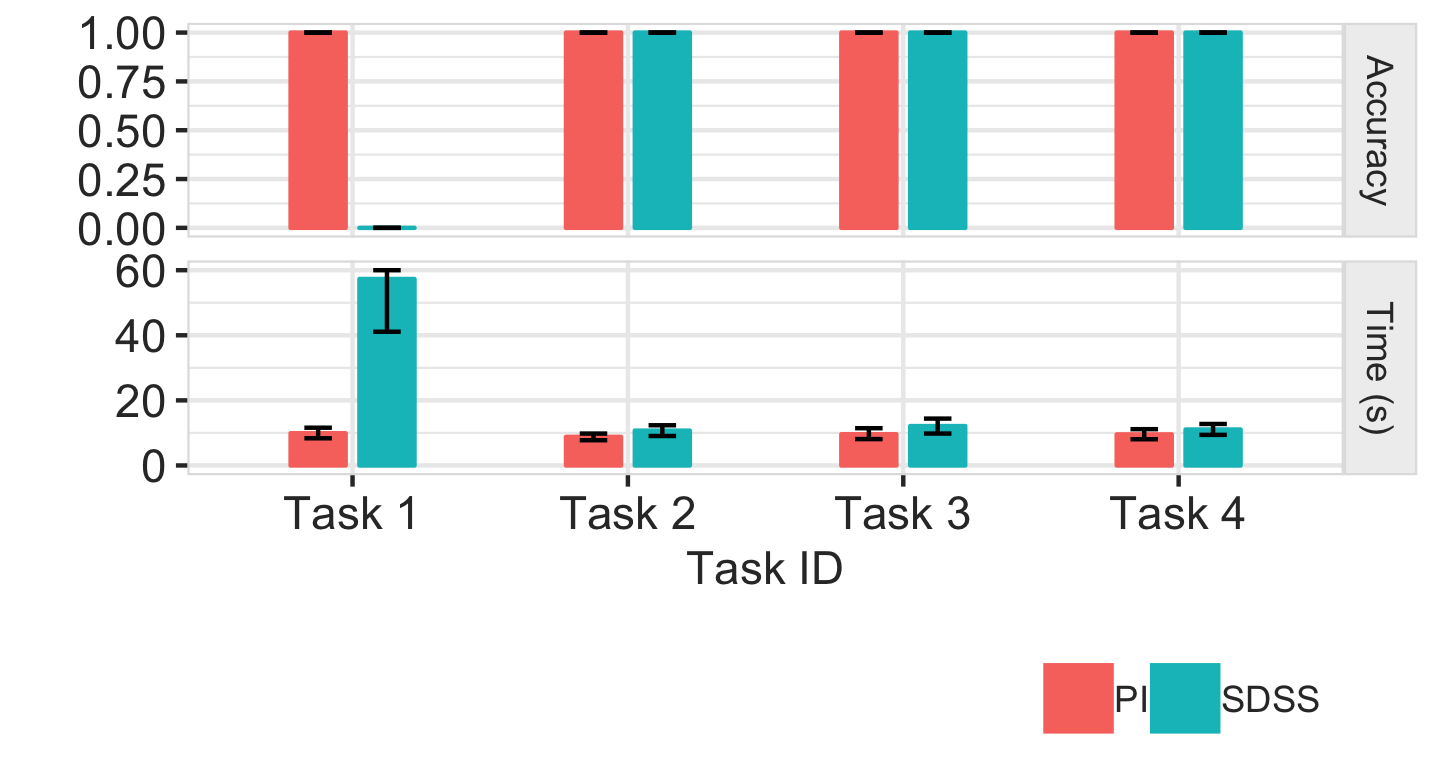}
    \caption{Time and accuracy using both interfaces.}
    \label{fig:userstudy}
  \end{subfigure}
  \caption{The original SDSS interface and the interface generated by \sys}
  \label{fig:allinterfaces}
\end{figure*}

\subsection{User Study}\label{s:userstudy}

We conducted a user study to understand how well \sys can identify and generate task-specific interfaces. We used the tiny SDSS query log sample~\cite{sdsslog} from 12/25-31/2003 and sampled 1000 queries to generate an interface (\Cref{fig:ui-2}).  These queries primarily perform 4 simple analysis tasks described in the SDSS manual~\cite{sdssmanual} of the Search Form Interface~\cite{sdssform}, which we call the ``SDSS interface''\footnote{From a pre-study, we re-styled the original SDSS interface (\Cref{fig:ui-sdss-orig}) to remove unnecessary widgets and styling.} (\Cref{fig:ui-1}).  Our goal is to understand the initial time to become acquainted with the interface, and the extent that \sys helps make specific tasks simpler to perform.   Detailed results, the original styled interface, and user feedback can be found in~\Cref{a:exp_user}.

We recruited 40 software engineers.  Each was randomly assigned the SDSS or generated interface, and asked to complete all 4 tasks in random order using the assigned interface. Task 1 finds objects with an \texttt{objectId}; Task 2 finds objects in a certain area;  Task 3 finds objects within a color range; Task 4 finds objects within a red-shift range.  Users were given 5 minutes to read the manual and examine  the assigned interface (in practice, they took around 2-3 minutes). We recorded the analysis times and accuracy of the first submission, and capped the time per task to $60$s.  For space reasons, the detailed results can be found in \Cref{a:exp_user}.

\Cref{fig:userstudy} depicts the average accuracy and time needed for each task under each condition, along with the $95^{th}$ confidence intervals. For Tasks 2-4, users submit their responses slightly faster using \sys ($9.3s\pm 0.8$, $95^{th}$ confidence interval) than using the SDSS interface ($11.2s\pm 1$).  In contrast, the times for Task 1 were respectively $9.9s\pm 1.5$, and $\approx 60$s.  This discrepancy is because the SDSS interface doesn't have dedicated widgets to lookup by objectid, and users need to manually write queries.  This highlights a benefit of \sys, which identified a task not supported by the SDSS interface, and created widgets specialized to it.  The task accuracies were identical for tasks 2-4.

We used the task, interface, and the task order (i.e., was this the $1^{st}$ task, $2^{nd}$, etc) as independent variables in an ANOVA test, where time was the dependent variable. All three variables are individually significant (p$\le2e^{-12}$ for all variables).  For instance, \Cref{a:exp_user} illustrates learning effects as participants complete more tasks, although the effects are not present for Task 1 using the SDSS interface.  The interaction between task and interface is also significant (p=$2e^{-16}$).    These results suggest that the generated interface plays a statistically significant role in how quickly users are able to complete the four tasks using \sys.

{\it \stitle{Takeaways: } The generated interface selected task-specific widgets as compared to the original SDSS interface.  It generated custom widgets for Task 1, which required manually writing queries in the original interface.  \sys was easier to initially use, and is slightly faster to complete tasks as compared to the original interface.}

\section{Conclusion and Discussion}\label{s:conclusion}

Interactive visual interfaces are increasingly relied upon in analysis and for end-users to interact with data.  However, knowing what analyses users want to perform is challenging, and creating interfaces for those analyses requires technical expertise. This paper proposed syntactic analysis of query logs for automatic interface generation, and a unified model to connect queries, changes between queries, interactive widgets, and interfaces.  We found that our approach is well-suited to analyses where query changes are structured and repeated, and less well-suited when there is unpredictable variation between queries (e.g., ad hoc analyses, or heterogeneous logs).  Our optimizations are able to generate interfaces for query logs with up to 10,000 queries within 10 seconds.

Stepping back, \sys is a ``quick and dirty'' approach towards custom analysis interface creation.  This data-driven approach does not work for analyses that have never been performed but that the user anticipates will be useful.  However, it is potentially ``good enough'' for a long tail of simple analyses.  Further expanding the scope of what analysis and settings \sys can scalably support can be viewed a progressive approach to this problem.

\stitle{Future Directions: } Using logs and data to generate analysis interfaces is a rich research direction. Interface quality can be improved by using metadata, language semantics, database content, as well as HCI user interface layout and design guidelines.   Multi-level interactions between widgets can leverage subtree co-occurrence statistics, and dependencies between queries can be identified as relationships between query results and subsequent queries.   Data cleaning can help distinguish queries from different tasks, anomalous queries, and different languages.  Finally, can our interface abstraction generate multi-modal applications, or bootstrap and enhance data science workflows such as Ava~\cite{John2017AvaFD}?

\begin{acks}
Thanks to Viraj Rai for earlier contributions, Anant Bhardwaj and Evan Jones for industry perspectives, reviewers for spot-on feedback, and NSF grants 1527765 \& 1564049.
\end{acks}

\clearpage
{
\bibliographystyle{abbrv}
\balance
\bibliography{main}
}

\appendix
\section{Cross-Client Experiment} \label{a:exp_recall_crossclient}

\begin{figure}[h!]
	\centering
	\includegraphics[width=.8\columnwidth]{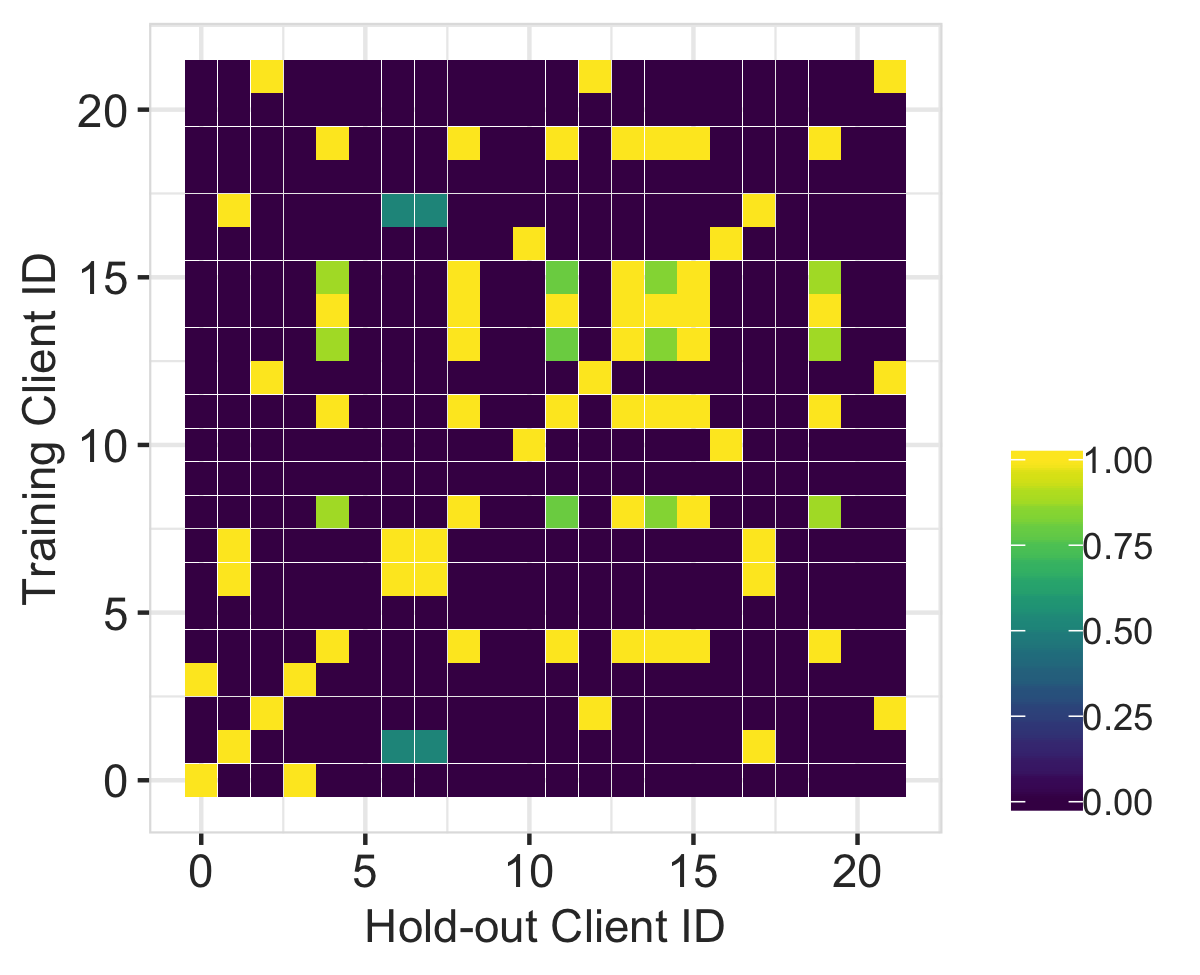}
	\caption{Pair-wise Recall Matrix.  Rows are training client IDs, cols are hold-out client IDs. }
	\label{fig:sdss_crossclient_matrix}
\end{figure}

\begin{figure}[h!]
	\centering
	\includegraphics[width=.8\columnwidth]{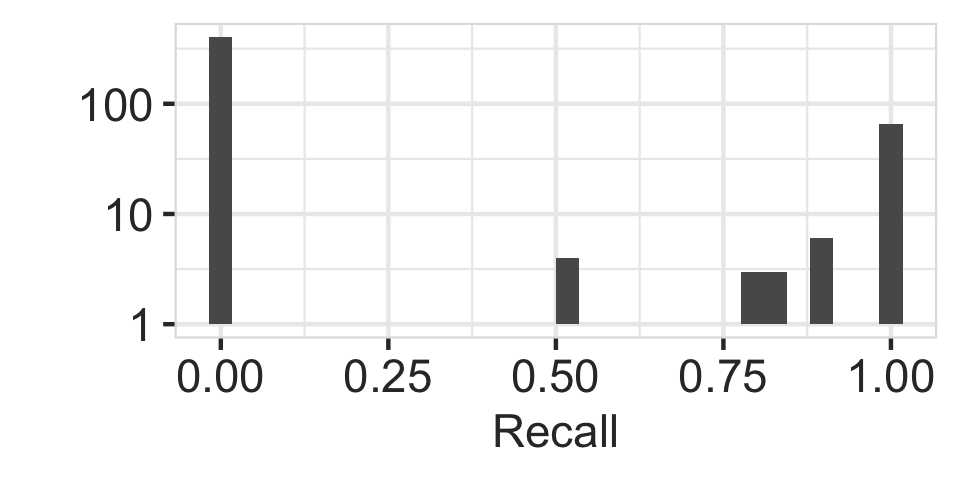}
	\caption{Histogram of hold-out recall.  Y-axis in log-scale.  }
	\label{fig:sdss_crossclient_recallhist}
\end{figure}

\review{\Cref{fig:sdss_crossclient_matrix} shows the pair-wise recall matrix for 22 random clients, where the value in row $i$ and column $j$ represents the recall of $Q_i$'s interface evaluated on $Q_j$.  We see in~\Cref{fig:sdss_crossclient_recallhist} that recall exhibits a bimodal distribution, where a given interface completely does not benefit hold-out client (recall=0), or it can fully express the hold-out client's queries (recall=1). }

\section{Complete Performance Experiments}\label{a:exp_perf}

We now evaluate the runtime performance of \sys as well as the effectiveness of the optimizations in \Cref{s:opt}.  We use the SDSS query log for these experiments.  In all of our experiments, the optimizations improve the runtime, but \textbf{do not affect the resulting interfaces}.  Thus, we focus solely on runtime.  We report the number of edges in the interaction graph, the interaction mining time, and the interface mapping time.

\begin{figure}[h]
    \includegraphics[width=\columnwidth]{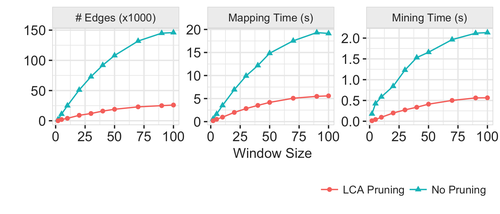}
    \vspace*{-.3in}
    \caption{Varying sliding window size and pruning optimizations for 100 queries.}
    \label{fig:time-prune}
\end{figure}

\stitle{Optimizations}
In this experiment, we use the per-client logs described in the recall experiments. We vary both the size of the sliding window (x-axis), as well as whether or not least common ancestor (LCA) pruning is used (lines).  The average query log size is 100 queries.

We see that LCA pruning dramatically reduces the size of the interaction graph---by as much as $5\times$ when the window size is 100 queries.   This naturally has a corresponding improvement in the interface mapping time, and a minor effect on the mining time because fewer edges need to be materialized.  This makes sense because interface mapping typically takes around $90\%$ of the total runtime, and reducing the number of edges considerably simplifies the problem size.   However, reducing the window size to 2 queries {\it reduces the total runtime to nearly zero}.  Note that the resulting interfaces remain the same.

\begin{figure}[h]
    \centering
    \includegraphics[width=\columnwidth]{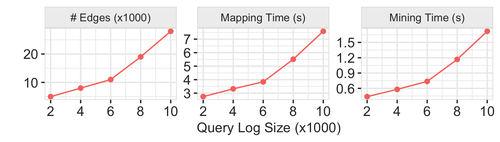}
    \vspace*{-.2in}
    \caption{Varying log size with window=2 and LCA pruning.}
    \label{fig:performance}
\end{figure}

\stitle{Scalability Experiments}
In this experiment, we use the full SDSS query log containing all client queries.  We do this to increase the total log size to 10,000 queries.  We set the sliding window size to 2, and enable LCA pruning.  
\Cref{fig:performance} shows that the number of edges and the runtime cost still increase quadratically with the log size.  Note that the number of edges is so low due to the small window size and pruning.  However, \sys is still able to generate the interactive interface within 10 seconds even with 10,000 queries in the log.

{\it \stitle{Takeaways: } For systematically changing query logs such as SDSS, \sys combines the sliding window and LCA pruning optimizations to reduce the end-to-end latency by multiple orders of magnitude. We expect that in practice, on logs of $\approx 2000$ queries, \sys runs in interactive time and generates interfaces within 3 seconds.}

\section{User Study}\label{a:exp_user}

\begin{figure}
	\centering
	\includegraphics[width=\columnwidth]{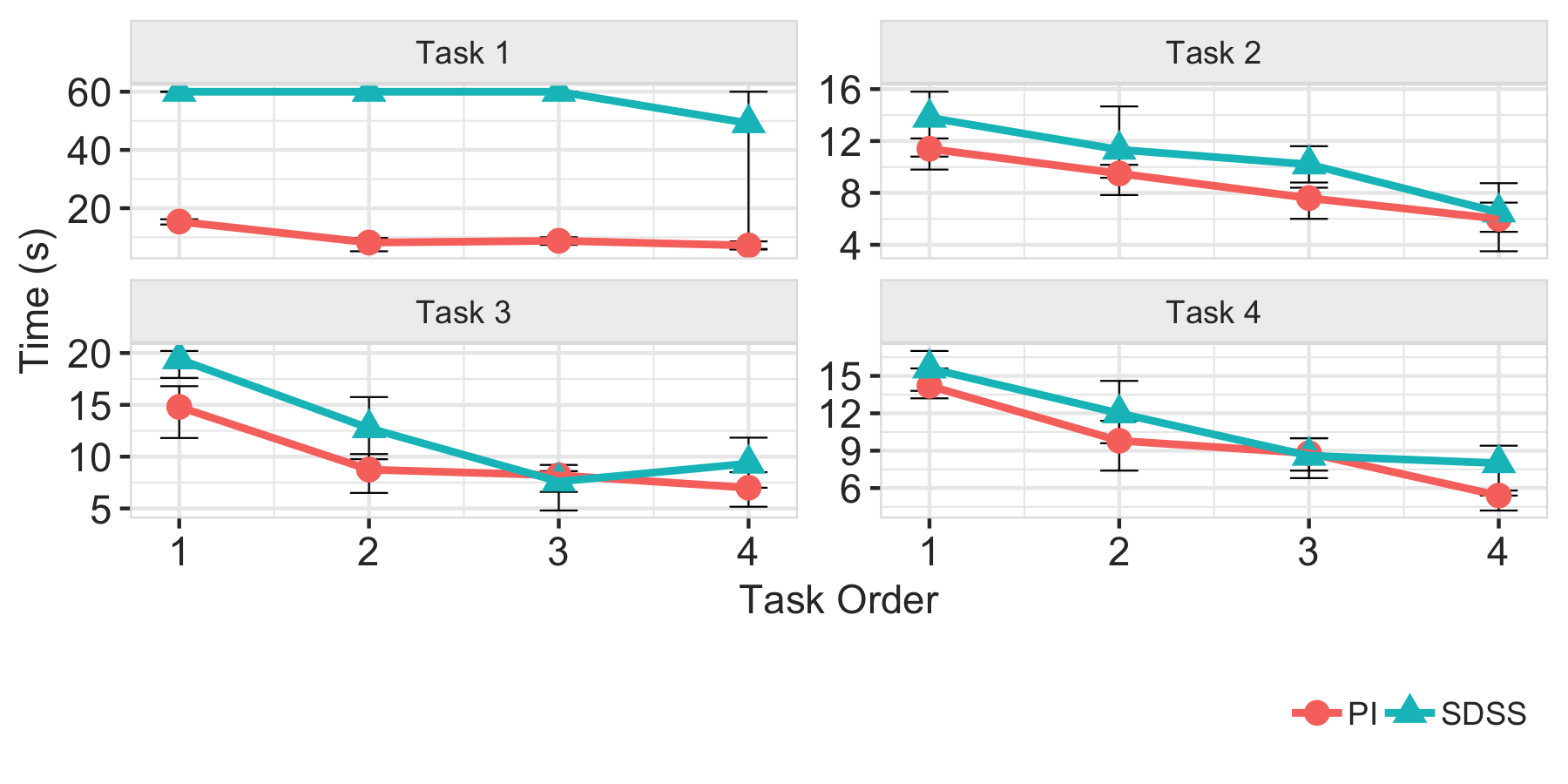}
    \caption{User study shows ordering effects (x-axis) on time for each task (facets) and assigned interface (lines). }
	\label{fig:user_time_orderpertask}
\end{figure}

\review{\Cref{fig:user_time_orderpertask} plots the time to complete each task as a function of the order (x-axis) that a given task (facet) was completed, and the interface (lines).  For instance, the upper left facet depicts the time to complete Task 1 when it is shown to the user as the first, second, third, or fourth task (x-axis).  We see that it takes more time to initially use the SDSS interface than the generated interface for all tasks (order=1), but users learn how to quickly use both interfaces as they complete more tasks.  \sys is considerably faster for Task 1 because the SDSS interface did not have widgets to perform the task, whereas \sys identified and generated widgets for those. }

\begin{figure}
	\centering
	\includegraphics[width=.8\columnwidth]{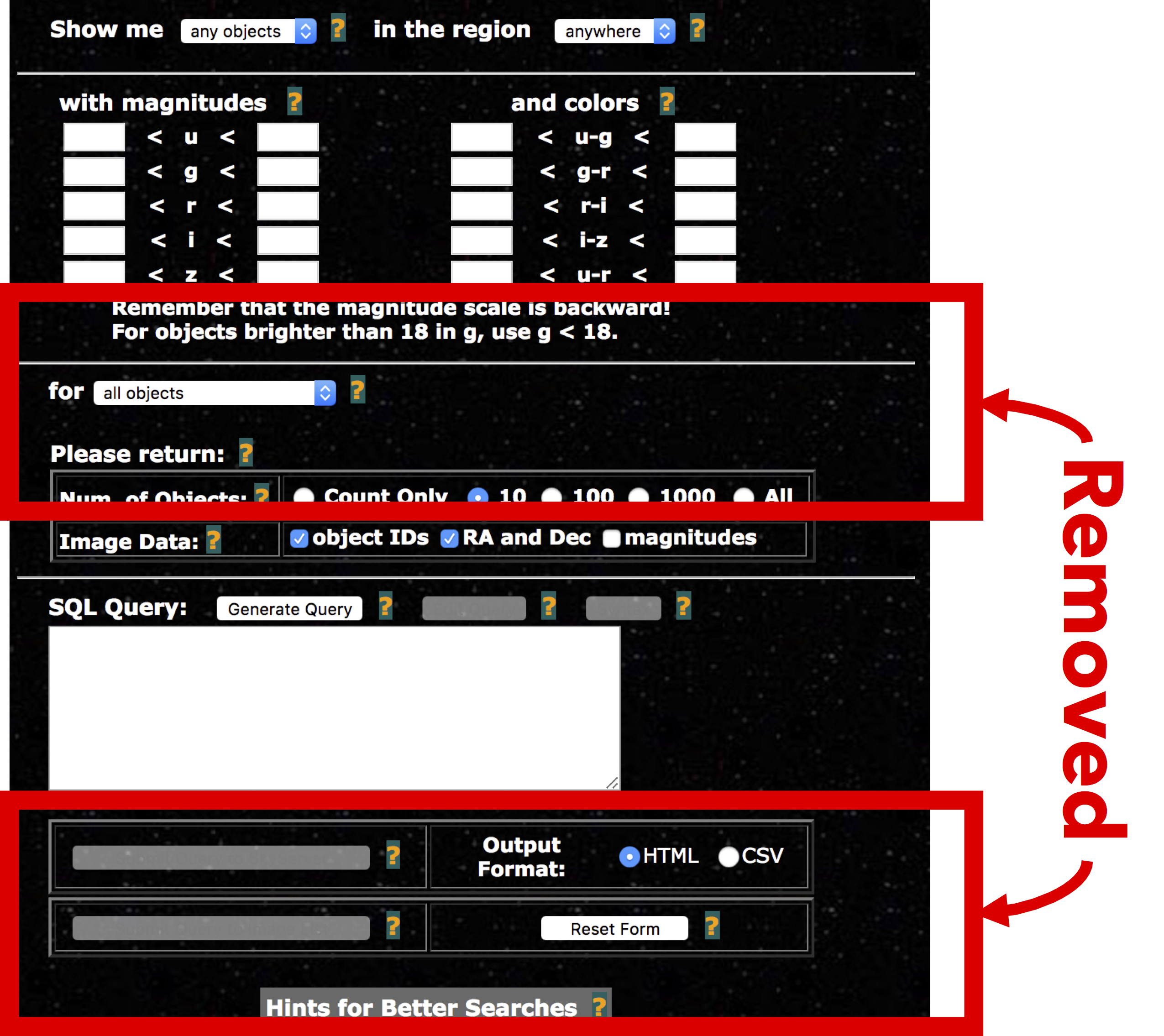}
	\caption{Original unstyled SDSS interface.  Red boxes were removed for user study.}
	\label{fig:ui-sdss-orig}
\end{figure}

\review{\Cref{fig:ui-sdss-orig} depicts the original SDSS interface with all widgets and existing styling.   The black background and superfluous widgets (e.g., the {\it Please Return: }, {\it Output format} controls) may artificially make it more difficult for participants.   We removed the styling and superfluous widgets (in the \red{red boxes}) to ensure an apples-to-apples comparison of the two user study interfaces. }

\review{Finally, we informally collected qualitative feedback from the users.  After users completed the tasks, we then showed and explained both interfaces to them, asked them for their preferred interface for the four tasks, and asked for general comments.  All candidates (irrespective of their assigned interface), preferred \sys over the SDSS interface.    We also summarize the positive and negative feedback below:}

\begin{itemize}[leftmargin=*]
  \item \review{{\it Confusing Widget Types:} Users found that the sliders were a bit confusing.   When the task asks the user to filter by an attribute range, then it is intuitive to leave text box widgets blank.  However, there isn't a default way to disable sliders to ensure they are not part of the query.    This suggests the value of providing better default widget presentations, and mechanisms to enable/disable groups of widgets.}
  \item \review{{\it Expertise and Fallbacks: } Users speculated that for expert SDSS users, the SDSS interface may be better because it has the ``fall back'' option of using the SQL textbox to write arbitrary SQL statements.  In cases where experts want to write an ad-hoc query, a textbox may be easier to use than a combination of slider/button/widgets.  }
  \item \review{{\it Keeping State:} Users mentioned that it would be great for the interface to remember previous state, instead of returning the widgets to their default values for every submission (since we reset the page for the next task).  }
  \item \review{{\it Simple Interactions: } Users liked that \sys did not have multi-level interactions, where the user needs to click a button for the desired widgets (e.g., text-boxes) to appear in the SDSS interface.  As first time users, they remarked that multi-level interactions are not very intuitive.}
\end{itemize}

\section{Precision Experiment} \label{a:exp_precision}

\begin{figure}[h!]
	\centering
	\includegraphics[width=.8\columnwidth]{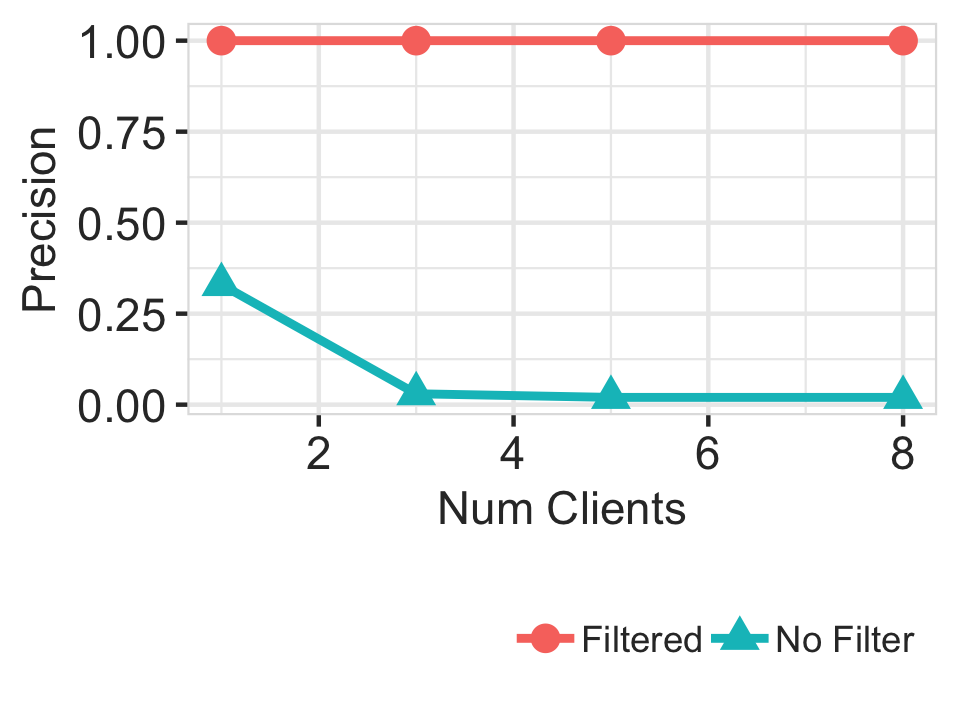}
	\caption{Results from Precision experiment}
	\label{fig:precision}
\end{figure}

\review{A purely syntactic approach to interface generate can easily generate queries that are nonsensical.  As a simple example, consider two widgets that respectively modify a table name in the \texttt{FROM} clause and an attribute name in the \texttt{WHERE} clause.  It is clear that picking an attribute from table T, but selecting table S in the \texttt{FROM} clause will result in an invalid query.  Many of the analyses in the SDSS log change the table, attribute, value, and more, thus this may be a prevalent issue.}

\review{To quantify this, we ran an experiment to evaluate the {\it Precision} of the generated interface's closure (\Cref{fig:precision}).  In other words, the percentage queries in the interface closure that do not violate the database schema.  To do so, we interleaved the same $M\in\{1, 3, 5, 8\}$ client logs as in the multi-client experiment (\Cref{s:exp_recall_multiclient}), and generated interfaces for each mixed log (x-axis).   We then created a local database with a schema consistent with the tables and attributes found in the queries---it ended up as a small subset of the SDSS database schema available online\footnote{\url{https://skyserver.sdss.org/dr12/en/help/browser/browser.aspx}}.  We then exhaustively enumerate the interface's closure (all queries it can express) and recorded the percentage of queries that can successfully run on the local database, and call that the {\it Precision}. Note that the database does not contain data, and we are not verifying the result.   }

\review{\Cref{fig:precision} shows that as we increase the heterogeneity of the input query log from $1$ to $8$, the precision drops from $\approx 30\%$ to around $1\%$ ({\it No Filter}).   In other words, $70-99\%$ of queries are rejected by the database. For this reason, our experiments used a simple filtering procedure to avoid generating such invalid queries---we keep a mapping from column name to the names of tables that contain the column in their schema, and verify that all column name node types have the containing table name node in the tree.  This procedure ({\it Filtered}) identifies queries that contain schema-related errors and increases precision to $100\%$.  In general, and as we discuss in future work, making use of the database schema as well as co-occurrence of subtrees in the query log can be an effective way of automatically avoiding invalid queries, and potentially inform better layout and interface quality.}

\end{document}